\pdfoutput=1
\PassOptionsToPackage{table,dvipsnames}{xcolor}
\documentclass{aa}
\usepackage{twoopt}
\usepackage{graphicx}
\usepackage{wrapfig}
\usepackage{float}
\usepackage{caption}
\usepackage{txfonts}
\usepackage{multirow}
\usepackage{tablefootnote}
\usepackage{amsmath}
\usepackage{indentfirst}
\usepackage{booktabs}

\usepackage{xcolor}
\definecolor{darkgoldenrod}{rgb}{0.72, 0.53, 0.04}

\usepackage[colorlinks = true,
  linkcolor = RoyalBlue,
  urlcolor = darkgoldenrod,
  citecolor = RoyalBlue,
  anchorcolor = RoyalBlue, 
  draft = false]{hyperref}

\makeatletter
\renewcommand*\aa@pageof{, page \thepage{} of \pageref*{LastPage}}
\makeatother

\usepackage{etoolbox}

\makeatletter
\patchcmd{\enddocument}
  {\let\@newl@bel\@testdef}
  {\let\@newl@bel\@testdef
   \NAT@swatrue
   \let\bibcite\NAT@testdef}
  {}
  {\PackageError{aa-fix}{Patch failed}{}}
\makeatother

\newcommand{\comment}[1]{}

\journalname{Preprint submitted to Astronomy \& Astrophysics}
\makeatletter
\renewcommand*\aa@manuscriptname{}
\renewcommand*\aa@textidlineempty{\textit{Preprint}}
\makeatother
\begin{document} 
   \title{Can integrated photometry and large-scale environment constrain the formation of polar rings in galaxies?}

  \author{O. V. Kompaniiets\inst{1} \corrauth{kompanietsolena@gmail.com} \and O. S. Pastoven\inst{1}\email{opastoven@mao.kiev.ua} \and D. V. Dobrycheva\inst{1}\email{dariadobrycheva@gmail.com} \and O. O. Hetmantsev\inst{1}\email{oleksandr.hetmantsev@gmail.com} \and I. B. Vavilova\inst{1}\email{irivav@mao.kiev.ua} \and I. O. Izviekova\inst{1,2}\email{izviekova@gmail.com} \and M. Yu. Vasylenko\inst{1}\email{vasmax@mao.kiev.ua}}

   \institute{
$^{1}$Main Astronomical Observatory of the NAS of Ukraine, 27, Akademik Zabolotny St., Kyiv, 03143, Ukraine\\
$^{2}$International Centre for Astronomical, Medical and Ecological Research, 27, Akademik Zabolotny St., Kyiv, 03143, Ukraine
}

\titlerunning{PRGs in Cosmic Web}
\authorrunning{Kompaniiets et al.}

   \date{}
\abstract
{Polar-ring galaxies (PRGs) contain stellar and gaseous structures
rotating in strongly misaligned planes. These structures may arise
from external gas acquisition through mergers, tidal accretion,
or filamentary supply.}
{We assess how far integrated photometry and the present-day environments of polar-ring galaxies can constrain ring formation scenarios. We examine whether the decline timescale of a secondary star-formation component is sufficiently well constrained to permit a comparison across filament environments.}
{We model GALEX, SDSS, and WISE photometry of nine PRGs with CIGALE
using two-component SFHs. Voronoi tessellation and DisPerSE
characterize environments using SDSS spectroscopic galaxies.
For seven targets at $z \geq 0.02$, we test 12 combinations of
reconstruction scale and persistence threshold. Filament association
accounts for reconstruction variability, filament widths inferred
from galaxy-density profiles, and estimated halo virial radii.}
{Star-formation rates span
$0.014$--$0.944\,M_\odot\,{\rm yr}^{-1}$.
Bayesian estimates of $\tau_2$ span $1.23$--$1.86$~Gyr,
with uncertainties of $1.21$--$1.41$~Gyr and strongly
overlapping intervals.
Missing far-infrared coverage of the thermal dust-emission peak
limits constraints on obscured star formation and contributes
to SFH uncertainty. SPRC056 is robustly filament-associated,
SPRC260 occupies a transition regime, and SPRC027, UGC7576,
and UGC7388 are robustly filament-distant. SPRC069 has a
reconstruction-sensitive geometry; SPRC179 remains distant
in all configurations but fails the formal spine-stability
criterion. The data do not establish a relation between
$\tau_2$ and filament proximity.}
{The reconstructed environments identify systems for which a filamentary supply channel is geometrically plausible at the observed epoch. The integrated photometry leaves the secondary-component timescale too uncertain to distinguish formation pathways. Spatially resolved measurements of the ring and host are needed to connect their stellar populations and gas kinematics to the inferred environment.}

   \maketitle
\nolinenumbers

\section{Introduction}

Polar-ring galaxies (PRGs) contain a central host and a
large-scale ring of gas, dust, and stars rotating in a strongly
misaligned, often nearly perpendicular, plane
\citep{Whitmore1990,Moiseev2011,Reshetnikov2019}. Their geometry
makes them useful systems for investigating the acquisition of
external material and the assembly of galactic components with
different angular momenta. Understanding how these structures
form and persist provides constraints on the roles of galaxy
interactions and gas accretion in galaxy evolution
\citep{Schweizer1983,Whitmore1987}.

The observational census of PRGs has expanded through systematic
searches in wide-field imaging surveys.
\citet{Dobrycheva2025} compiled the Catalogue of Inspected Polar
Ring Galaxies (CIPRG), comprising 179 visually inspected objects,
including seven newly identified candidates
\citep{Dobrycheva2025cat}. The catalogue combines objects from
previous PRG catalogues with new candidates identified through
deep learning and visual inspection of SDSS images. Five further
strong PRG candidates were identified during the expert validation
of CNN-based SDSS morphological catalogues \citep{Dobrycheva2026}.
A recent compilation by \citet{Yu2026} includes 40 kinematically
confirmed PRGs.

The principal proposed formation channels involve mergers,
tidal transfer of material from neighboring galaxies, and
accretion from the surrounding large-scale environment.
Simulations show that mergers and tidal encounters can produce
polar configurations through the redistribution of gas and stars
\citep{Bekki1997,Bournaud2003,Quiroga2020}. Observational studies of ESO~287-IG50, NGC~4111, and NGC~3808A/B provide examples in which interactions or companion capture have been proposed \citep{Freitas2024,Roier2022,Ordenes2016}. In Illustris TNG50, \citet{Smirnov2024} identified six PRG analogues whose polar rings formed through interactions with companions, either through star formation in accreted gas or through tidal disruption of a satellite's stellar component. The variety of these pathways motivates the use of additional diagnostics beyond polar morphology.

Accretion from the Cosmic Web provides another possible source
of material with angular momentum misaligned with that of the
host. Cosmological simulations by \citet{Maccio2006} demonstrated the formation of a polar ring from cold gas supplied along a filament, while \citet{Brook2008} showed that continued gas accretion following the assembly of an inner disk can produce a long-lived, highly inclined star-forming disk. More generally, external supply can replenish a gas reservoir and alter its angular momentum. \citet{Taylor2018} linked counter-rotating gaseous disks to gas acquisition as galaxies enter filaments, and \citet{vandeVoort2015} showed that continued accretion can sustain misalignment between gaseous and stellar components. These studies motivate examining both the subsequent stellar growth and the environmental setting of externally supplied galactic structures.

Chemical abundances and resolved kinematics provide complementary constraints on these pathways. The low gas-phase metallicity and approximately flat abundance gradient of the polar disk in NGC~4650A are consistent with the acquisition of relatively metal-poor material \citep{Spavone2010}, while counter-rotation between its inner and outer regions indicates a complex angular-momentum history \citep{Coccato2014}. Metallicity measurements in UGC~7576 and UGC~9796, including the absence of a significant abundance gradient in UGC~7576, have likewise been used to assess accretion and interaction scenarios \citep{Spavone2011}. Studies of VGS31b and VGS~12 connect inclined or polar gas structures with their larger-scale
surroundings \citep{SpavoneIodice2013,Stanonik2009,Egorova2026}.

Gas reservoirs and stellar populations provide information on
the subsequent growth of these systems. Molecular-gas
observations have revealed substantial reservoirs in several
PRGs \citep{Combes2013}. Multiwavelength spectral energy
distribution (SED) modeling has shown that polar components
can contain stellar masses comparable to those of their hosts
\citep{Lackey2024}. Comparisons of NGC~3718, NGC~2685, and
NGC~4262 further illustrate the diversity of stellar-population
ages, star-formation rates, and gas content among PRGs
\citep{Akhil2025}. Such measurements motivate the use of
star-formation histories (SFHs) to investigate the temporal
development of stellar populations associated with external
assembly.

Environment can influence both the formation and the survival of
polar rings, as close companions are known to affect the morphology
and colours of galaxies \citep{Melnyk2012}. Using projected distances
to the nearest companion and companion counts, \citet{SavchenkoReshetnikov2017}
found that PRG candidates inhabit, on average, less dense environments
than comparison galaxies on scales of hundreds of kiloparsecs. Reduced
perturbations may favor the survival of polar rings after
their formation. Local galaxy density and filament geometry
therefore provide complementary environmental information:
the former characterizes the immediate galaxy distribution,
while the latter places a system within the larger-scale
network potentially associated with gas supply.

Building on our two-component SFH analysis of SDSS~J140644.42+471602.0 \citep{Dobrycheva2025}, we ask what integrated photometry and the present-day large-scale environment can tell us about the formation of polar rings. We model the integrated SEDs of nine galaxies from the CIPRG and reconstruct the filament environments of seven. We first assess whether the photometry constrains the decline timescale of a secondary star-formation component sufficiently well to distinguish individual systems. We then test which galaxies lie close to a reconstructed filament and how sensitive that inference is to the reconstruction. We use these results to assess whether differences in the inferred SFHs can be interpreted in the context of present-day filament proximity.

Throughout this work, we adopt the Planck 2018 cosmology
\citep{Planck2020}, with $H_0=67.66~{\rm km\,s^{-1}\,Mpc^{-1}}$,
$\Omega_{\rm m}=0.3097$, and $\Omega_{\Lambda}=0.6889$.
All comoving distances and cosmology-dependent quantities are
calculated using this cosmology.

The dataset and sample selection are described in Section~\ref{data}. Section~\ref{multiwave} presents the photometric measurements, SED modeling, and environmental methods, including the Voronoi and DisPerSE analyses in Section~\ref{sec:cosmic_web}. The results and their implications for PRG formation pathways are discussed in Section~\ref{discuss}, and the conclusions are summarized in Section~\ref{conclusion}.
\section{Dataset} \label{data}
\subsection{Available Data}
In our previous work we compiled a catalog of PRGs -- CIPRGs \citep{Dobrycheva2025cat} -- from available literature with addition of newly discovered PRGs by \cite{Dobrycheva2025}. The CIPRGs features 167 objects from catalogs by \cite{Whitmore1990}, \cite{Moiseev2011}, and \cite{Reshetnikov2019}, which were visually inspected and assigned 'strong', 'good', and 'weak' categories (objects showing no resemblance to a polar structure
were not included). The CIPRGs also features 5 objects found in the works of \cite{Skryabina2024}, \cite{Akhil2024}, \cite{Freitas2024}, and \cite{Stanonik2009}, as well as 7 objects found by \cite{Dobrycheva2025}. For this study, 26 objects from the 'strong' category were selected.

\subsection{Sample selection}
Our parent sample comprises 26 strong PRG candidates. For the multiwavelength analysis, we selected galaxies with identifiable source emission suitable for aperture photometry in the SDSS optical and WISE infrared images. GALEX ultraviolet measurements were included where available; a UV detection was not required for inclusion in the sample. For the present study, we selected nine polar-ring galaxies
from the parent sample for integrated photometric analysis.We additionally searched for deeper or complementary observations at mid- and far-infrared wavelengths, including measurements capable of sampling the thermal dust-emission peak, as well as for publicly available radio observations. No such additional infrared data were available for the selected galaxies. The infrared coverage used in this work is limited to the WISE bands, with no far-infrared photometry. The dataset includes all publicly accessible observations currently available for the multi-wavelength analysis of these galaxies.

\section{Methods} 
\label{multiwave}
\subsection{Aperture Photometry}
We performed aperture photometry for the nine selected galaxies using SDSS optical and WISE infrared images, supplemented by GALEX ultraviolet images where available.The GALEX images were downloaded from the Mikulski Archive for Space Telescopes (MAST), the SDSS images were taken from the SDSS Science Archive Server, and the WISE images were taken from the IRSA archive. 
The galaxies in our sample have angular diameters on the order of a minute. We performed aperture photometry with the Aperture Photometry Tool
(APT), following the procedure described in our previous study \citep{Pastoven2024, Kompaniiets2026}.
Because the detectable extent of each galaxy varies with wavelength,
we used the radial surface-brightness profile and the curve of growth
to select the source aperture and local background region. We chose
the aperture where further enlargement produced no appreciable
increase in the background-subtracted flux. The adopted apertures are listed in Tables~1--3. We used the sky-average model in APT
to subtract the local background. Sky average subtraction model. The available background-subtraction models are described by \citet{Laher2012}.
The photometric fluxes in the GALEX (FUV, NUV), SDSS ($ugriz$),
and WISE ($W1$--$W4$) bands were calculated from the AB magnitudes
following the procedure described in \citet{Kompaniiets2026}.

For GALEX, we used the magnitudes in the FUV and NUV filters
with zero-point magnitudes $m_{0,\mathrm{FUV}} = 18.82$ and
$m_{0,\mathrm{NUV}} = 20.08$, which correspond to the AB system
(see Table~\ref{tab:FluxesUV}).

For the SDSS ($u$, $g$, $r$, $i$, $z$), we used the universal zero magnitude $m_0 = 22.5$, which corresponds to the AB system (see Table \ref{tab:Fluxesoptic}).

For processing of the WISE photometric data ($W1$-$W4$), instrumental zero-point magnitudes $m_0$ were initially used to calculate magnitudes in the Vega system.
The values of $m_0$ are taken from the official WISE documentation:\\
$m_{0,W1}=20.5$,\\
$m_{0,W2}=19.5$,\\
$m_{0,W3}=18.0$,\\
$m_{0,W4}=13.0.$\\
\\
After calculating the magnitudes in the Vega system, they were converted to the AB system by adding standard offsets (see Table \ref{tab:FluxesIR}):
\\
$m_{AB(W1)}=m_{Vega(W1)}+2.699$,\\
$m_{AB(W2)}=m_{Vega(W2)}+3.339$,\\
$m_{AB(W3)}=m_{Vega(W3)}+5.174$,\\
$m_{AB(W4)}=m_{Vega(W4)}+6.620$.\\
\\
The UV-to-IR photometry was corrected for foreground Galactic
reddening and extinction using the line-of-sight extinction values
provided by the IRSA Dust Extinction Service, following the same
procedure as in \citet{Dobrycheva2025}.

Blank entries in Tables~1--3 correspond to non-detections in the
respective bands. Where an upper limit could be estimated, it was included
in the CIGALE fitting using the \texttt{noscaling} upper-limit treatment;
otherwise, the corresponding band was excluded from the fit.

\begin{table*}
\small
\centering
\caption{Aperture photometry fluxes in the UV bands. Flux densities are given in mJy.}
\label{tab:FluxesUV}
\begin{tabular}{|c|c|c|c|c|c|}
\hline
Name             & RA       & DEC     & FUV              & NUV      & Aperture size        \\ \hline
IC1689           & 20.9494  & 33.0553 & 0.04 $\pm$ 0.06  & 0.08 $\pm$ 0.05 &  23" $\times$ 20"\\
SPRC027          & 176.1834 & 23.1625 &                  & 0.04 $\pm$ 0.04 & 45" $\times$ 18"\\
SPRC069          & 312.0236 & 0.0688  & 0.01 $\pm$  0.04 & 0.03 $\pm$ 0.03 & 23" $\times$ 8"\\
SPRC056          & 227.8087 & 37.0438 & 0.07 $\pm$  0.09 & 0.10 $\pm$  0.06 & 44" $\times$ 30"\\
UGC7576          & 186.9244 & 28.6981 &                  & 0.10 $\pm$ 0.06  & 41" $\times$ 21"\\
SPRC260 & 176.3761 & 9.7291  & 0.36 $\pm$  0.20 & 0.38 $\pm$  0.12 & 38" $\times$ 21"\\
SPRC241 & 40.4875  & -6.7923 &                  & 0.21 $\pm$ 0.09  & 45" $\times$ 30"\\
UGC7388 & 185.0656 & 33.6610 & 0.04 $\pm$ 0.04  & 0.44 $\pm$  0.22 & 53" $\times$ 27"\\
SPRC179 & 255.7866 & 46.3736 & 0.07 $\pm$  0.04 & 0.86 $\pm$  0.31 & 32" $\times$ 23"\\\hline

\end{tabular}%

\end{table*}

\begin{table*}
\small
\centering
\caption{Aperture photometry fluxes in the Optical bands. Flux densities are given in mJy.}
\label{tab:Fluxesoptic}
\begin{tabular}{|c|c|c|c|c|c|c|c|c|}
\hline
Name    & RA       & DEC     & u                 & g                 & r                 & i                 & z    &       Aperture size    \\ \hline
IC1689  & 20.9494  & 33.0553 & 0.95 $\pm$  0.08  & 4.15  $\pm$  0.13 & 8.72  $\pm$  0.18 & 12.79 $\pm$ 0.22  & 16.47 $\pm$  0.31 & 24" $\times$ 11" \\
SPRC027 & 176.1834 & 23.1625 & 0.17  $\pm$  0.03 & 0.51 $\pm$  0.04  & 1.05 $\pm$  0.06  & 1.52  $\pm$  0.08 & 2.11  $\pm$  0.10 & 30" $\times$ 13"\\
SPRC069 & 312.0236 & 0.0688  & 0.36  $\pm$  0.01 & 1.46 $\pm$  0.14  & 2.98  $\pm$  0.11 & 4.37  $\pm$ 0.23  & 5.66  $\pm$  0.09 & 16" $\times$ 8"\\
SPRC056 & 227.8087 & 37.0438 & 0.45  $\pm$ 0.05  & 2.01 $\pm$  0.09  & 4.20  $\pm$  0.19 & 6.23  $\pm$  0.15 & 8.28  $\pm$  0.20 & 44" $\times$ 28"\\
UGC7576 & 186.9244 & 28.6981 & 0.52  $\pm$ 0.09  & 2.67  $\pm$ 0.10  & 5.24  $\pm$  0.15 & 7.76  $\pm$  0.19 & 9.03 $\pm$  0.38  & 48" $\times$ 18"\\
SPRC260 & 176.3761 & 9.7291  & 0.75  $\pm$  0.07 & 3.04 $\pm$  0.11  & 5.85 $\pm$  0.15  & 8.28  $\pm$  0.18 & 10.21 $\pm$  0.30 & 36" $\times$ 23"\\
SPRC241 & 40.4875  & -6.7923 & 0.83  $\pm$  0.07 & 4.07 $\pm$ 0.12   & 7.89  $\pm$  0.17 & 11.29 $\pm$  0.21 & 14.56 $\pm$  0.27 & 46" $\times$ 20"\\
UGC7388 & 185.0656 & 33.6610 & 1.26  $\pm$  0.15 & 3.24 $\pm$  0.30  & 6.10  $\pm$  0.16 & 8.96  $\pm$  0.20 & 11.88 $\pm$  0.38 & 51" $\times$ 31"\\
SPRC179 & 255.7866 & 46.3736 & 1.02 $\pm$  0.08 & 4.52 $\pm$  0.13  & 8.78  $\pm$  0.18 & 12.32 $\pm$  0.22 & 15.48 $\pm$  0.36 &  30" $\times$ 23"\\ \hline
\end{tabular}
\end{table*}

\begin{table*}
\small
\centering
\caption{Aperture photometry fluxes in the IR bands. Flux densities are given in mJy.}
\label{tab:FluxesIR}

\begin{tabular}{|c|c|c|c|c|c|c|c|}
\hline
Name    & RA       & DEC     & W1     & W2   & W3          & W4 & Aperture size    \\ \hline
IC1689  & 20.9494  & 33.0553 & 12,06  $\pm$  0.15 & 6.65  $\pm$  0.135 & 7.12  $\pm$  0.122 & 4.09  $\pm$  0.46 & 25" $\times$ 14"\\
SPRC027 & 176.1834 & 23.1625 & 1.53  $\pm$  0.05 & 0.91  $\pm$  0.05 & 0.77  $\pm$  0.04 &             &     15" $\times$ 15" \\
SPRC069 & 312.0236 & 0.0688  & 4.03  $\pm$  0.09 & 2.48  $\pm$  0.08 & 11.87  $\pm$  0.19 & 5.89  $\pm$ 0.54  & 17" $\times$ 17"\\
SPRC056 & 227.8087 & 37.0438 & 5.01  $\pm$  0.08 & 2.86  $\pm$  0.08 & 2.28  $\pm$  0.06 & 1.35  $\pm$  0.27 & 22" $\times$ 17"\\
UGC7576 & 186.9244 & 28.6981 & 6.72  $\pm$  0.12 & 3.75  $\pm$  0.11 & 2.77  $\pm$  0.10 &                  & 41" $\times$ 21" \\
SPRC260 & 176.3761 & 9.7291  & 7.37  $\pm$  0.13 & 4.04  $\pm$  0.11 & 2.59  $\pm$  0.12 &                   & 32" $\times$ 22" \\
SPRC241 & 40.4875  & -6.7923 & 9.1  $\pm$  0.26 & 5.07  $\pm$  0.2 & 2.53  $\pm$  0.11 &                   & 48" $\times$ 28"\\
UGC7388 & 185.0656 & 33.6610 & 13.40 $\pm$  0.39 & 9.04  $\pm$  0.36 & 44.46 $\pm$  0.31 & 61.45 $\pm$  1.81 & 41" $\times$ 34"\\
SPRC179 & 255.7866 & 46.3736 & 11.91  $\pm$  0.16 & 7.02 $\pm$  0.14  & 12.38 $\pm$  0.17 & 9.42  $\pm$  0.93 & 34" $\times$ 28" \\ \hline
\end{tabular}
%
\end{table*}

\subsection{SED modeling with CIGALE}
\label{sec:cigale}

The physical properties of the PRGs were inferred by modeling their integrated UV-to-IR spectral energy distributions with CIGALE \citep{Boquien2019}. The fitted photometry comprises the GALEX FUV and NUV bands, the SDSS $u$, $g$, $r$, $i$, and $z$ bands, and the four WISE bands W1--W4. The physical quantities inferred from the SED fitting depend on the adopted star-formation history, stellar-population models, attenuation prescription, dust-emission templates, and explored parameter grid.

We adopted the \texttt{sfh2exp} module as a phenomenological
representation of a main stellar population and a later episode
of stellar-mass growth. The SFH consists of two exponentially
declining components. The main component is characterized by
an e-folding timescale $\tau_{\rm main}$ and age $t_{\rm main}$,
while the secondary component is described by
$\tau_{\rm burst}$, $t_{\rm burst}$, and a stellar-mass fraction
$f_{\rm burst}$. Throughout this work, we use
$\tau_2 \equiv \tau_{\rm burst}$ to denote the decline timescale
of the secondary component.

This parameterization provides a common basis for comparing
the integrated SFHs of the sample galaxies with their
reconstructed environments. A secondary component is included
in all models, so its presence is an assumption of the adopted
model family. The analysis constrains its parameters within
that family and does not establish that two distinct episodes
are required by the photometry. The fitted $\tau_2$ describes
an exponential decline and is distinct from the episode age
and the duration of external gas supply. Since the photometry
combines the host and polar structure, the temporal components
do not spatially separate these structures.

The stellar emission was computed using the \texttt{bc03} stellar-population synthesis models of \citet{Bruzual2003}, assuming a Chabrier initial mass function \citep{Chabrier2003} and solar metallicity, $Z=0.02$. Nebular line and continuum emission were included with the \texttt{nebular} module following \citet{Inoue2011}. We explored ionization parameters $\log U=-4$, $-3$, and $-2$ and gas metallicities $Z_{\rm gas}=0.001$, 0.020, 0.030, and 0.051, while fixing the electron density to $100~{\rm cm}^{-3}$. The escape fraction and the fraction of Lyman-continuum photons absorbed by dust were both fixed to zero.

Dust attenuation was modeled with the modified Charlot \& Fall two-component prescription implemented in \texttt{dustatt\_modified\_CF00} \citep{Charlot2000}. Dust emission was described using the \texttt{dale2014} templates \citep{Dale2014}. The dust-heating parameter was fixed to $\alpha=2.0$, while the fractional AGN contribution was allowed to vary between 0 and 0.5. The \texttt{redshifting} module was applied using the individual redshifts provided in the input data file.

The complete CIGALE parameter grid is given in Table~\ref{tab:param}. The varying parameters correspond to 622,080 model combinations at each model redshift. A relative systematic uncertainty of 10\% was added in quadrature to the photometric uncertainties. Upper limits, when present, were included using the \texttt{noscaling} treatment implemented in CIGALE.

We assessed the recovery of the physical parameters using the mock analysis implemented in CIGALE. For each galaxy, CIGALE constructs a synthetic observation from the fitted model, perturbs the model fluxes according to the observational uncertainties, and analyses the resulting mock SED using the same model grid. The parameters recovered from the mock catalog can then be compared with their known input values to assess how well the adopted photometric coverage and model grid constrain the quantities used in the analysis. We use this test to evaluate the recovery of the stellar mass, star-formation rate, and parameters describing the secondary star-formation component. The recovery tests show that the SFR, stellar mass, and attenuation parameters are well recovered, whereas the detailed SFH timescales and stellar-population ages are considerably more degenerate. The complete mock-analysis results are presented in Appendix~\ref{app:mock}. Given the comparatively weak recovery of $\tau_{2}$, we use this parameter only to search for broad differences among the individual systems and do not interpret small differences between their Bayesian central estimates as physically resolved.

The available wavelength coverage limits the physical
interpretation of the fitted SFHs. The WISE bands constrain
the near- and mid-infrared emission, but no far-infrared
measurements sample the thermal dust-emission peak.
Consequently, the total dust-reprocessed luminosity and the
obscured contribution to star formation are less directly
constrained \citep{Boquien2019}. The low signal-to-noise ratios
of many GALEX measurements further limit the constraints on
recent stellar growth. Degeneracies among the secondary
component's age, mass fraction, decline timescale, and dust
attenuation also affect SFH recovery from integrated
broad-band photometry \citep{Conroy2013}.

The inferred parameters are additionally conditional on the
fixed stellar metallicity, attenuation-law slopes, and
dust-heating parameter. These model choices are not varied
in the mock analysis. Broader photometric coverage would
strengthen the observational constraints, while spatially
resolved spectroscopy would provide complementary tests of
the stellar-population interpretation.

\begin{table*}
\centering  
\caption{Input parameters of the CIGALE model grid.}
\label{tab:param}
\begin{tabular}{lc}\hline \hline

\multicolumn{2}{c}{\texttt{sfh2exp}} \\ \hline           
e-folding time of the main stellar population model in Myr.
   & 500, 1000, 2000 \\ 
e-folding time of the late starburst population  model in Myr.
& 500, 1000, 3000, 4000 \\ 
  Mass fraction of the late burst population. & 0.001, 0.01, 0.05, 0.1, 0.2, 0.3 \\ 

   Age of the main stellar population in the galaxy in Myr.  & 5000, 7000, 9000, 11000 \\        
   Age of the late burst in Myr.  &  100, 500, 1000, 2000, 3000, 4000  \\\hline 
   
\multicolumn{2}{c}{\texttt{bc03}} \\ \hline
Initial mass function & \cite{Chabrier2003}\\ 
Metalicity & 0.02 \\
Age [Myr] of the separation between the young and the old star populations         &     10     \\ \hline
\multicolumn{2}{c}{\texttt{nebular}} \\\hline

Ionization parameter. & -2.0, -3.0, -4.0\\
Gas metallicity. &  0.001, 0.02, 0.03, 0.051\\ 
Electron density. & 100\\ 
Line width in km/s & 300\\\hline 
\multicolumn{2}{c}{\texttt{dustatt\_modified\_CF00}} \\\hline

V-band attenuation in the interstellar medium. & 0.05, 0.1, 0.25, 0.5, 0.75, 1.0  \\ 
$Av_{ISM} / (Av_{BC}+Av_{ISM})$ & 0.44\\ 
    
    Power law slope of the attenuation in the ISM. &  -0.7 \\
    Power law slope of the attenuation in the birth clouds. &  -1.3\\\hline 
\multicolumn{2}{c}{\texttt{dale2014}} \\ \hline
 AGN fraction & 0.0, 0.05, 0.1, 0.25, 0.5 \\ 
 $\alpha$ & 2.0\\\hline \hline
\end{tabular}
\end{table*}

\begin{table*}
\centering
\small
\setlength{\tabcolsep}{2.8pt}
\caption{CIGALE results.}
\label{tab:res}

\begin{tabular}{lccccccccc}
\hline
\hline
ID &
$A_{V,\rm BC}$ &
$A_{V,\rm ISM}$ &
$A_{\rm FUV}$ &
$A_{V,\rm B90}$ &
SFR &
$M_{\star}$ &
$L_{\star}$ &
$\tau_2$ &
$\chi^2_{\nu}$ \\
&
&
&
&
&
($M_{\odot}\,{\rm yr}^{-1}$) &
($10^{9}M_{\odot}$) &
($10^{9}L_{\odot}$) &
(Gyr) &
\\
\hline

IC1689
& $0.12\pm0.02$
& $0.09\pm0.02$
& $0.44\pm0.11$
& $0.09\pm0.02$
& $0.014\pm0.012$
& $16.66\pm2.66$
& $12.92\pm0.65$
& $1.34\pm1.26$
& 1.73 \\

SPRC027
& $0.10\pm0.03$
& $0.08\pm0.02$
& $0.45\pm0.13$
& $0.08\pm0.02$
& $0.179\pm0.105$
& $19.85\pm3.70$
& $18.06\pm0.90$
& $1.74\pm1.37$
& 0.23 \\

SPRC069
& $0.32\pm0.01$
& $0.25\pm0.01$
& $1.26\pm0.05$
& $0.25\pm0.01$
& $0.145\pm0.048$
& $14.61\pm2.49$
& $13.96\pm0.70$
& $1.68\pm1.34$
& 4.11 \\

SPRC056
& $0.06\pm0.00$
& $0.05\pm0.00$
& $0.30\pm0.02$
& $0.05\pm0.00$
& $0.466\pm0.212$
& $89.50\pm13.59$
& $81.91\pm4.10$
& $1.56\pm1.33$
& 1.19 \\

UGC7576
& $0.11\pm0.03$
& $0.09\pm0.02$
& $0.45\pm0.12$
& $0.09\pm0.02$
& $0.053\pm0.041$
& $21.42\pm3.23$
& $17.82\pm0.89$
& $1.55\pm1.34$
& 0.31 \\

SPRC260
& $0.06\pm0.00$
& $0.05\pm0.00$
& $0.30\pm0.01$
& $0.05\pm0.00$
& $0.179\pm0.048$
& $17.51\pm3.14$
& $16.95\pm0.85$
& $1.59\pm1.30$
& 0.33 \\

SPRC241
& $0.06\pm0.00$
& $0.05\pm0.00$
& $0.28\pm0.03$
& $0.05\pm0.00$
& $0.036\pm0.024$
& $16.01\pm1.99$
& $14.23\pm0.71$
& $1.46\pm1.31$
& 0.60 \\

UGC7388
& $1.08\pm0.16$
& $0.85\pm0.12$
& $3.01\pm0.29$
& $0.87\pm0.12$
& $0.944\pm0.289$
& $33.13\pm5.94$
& $42.82\pm2.14$
& $1.86\pm1.41$
& 1.20 \\

SPRC179
& $0.30\pm0.06$
& $0.23\pm0.05$
& $1.07\pm0.22$
& $0.24\pm0.05$
& $0.130\pm0.092$
& $67.59\pm7.97$
& $63.02\pm3.15$
& $1.23\pm1.21$
& 2.08 \\

\hline
\end{tabular}

\tablefoot{
$A_{V,\rm B90}$ is the effective rest-frame V-band attenuation
computed by CIGALE using the \texttt{V\_B90} filter.
All attenuation values are expressed in magnitudes.
$\tau_2$ is the Bayesian estimate of the e-folding timescale
of the secondary star-formation component in the adopted
two-component SFH model.
$\chi^2_{\nu}$ is the reduced chi-square of the best-fit
CIGALE model.
The quoted uncertainties are the standard deviations of
the likelihood-weighted parameter distributions within
the adopted model grid.
Values rounded to 0.00 indicate formal uncertainties below
the displayed precision and do not imply zero physical
uncertainty.
Small formal uncertainties may arise when the likelihood
is concentrated on a single grid value; uncertainties
associated with the fixed model assumptions are not included.
}
\end{table*}

\begin{figure*}
    \centering
    \includegraphics[width=0.45\linewidth]{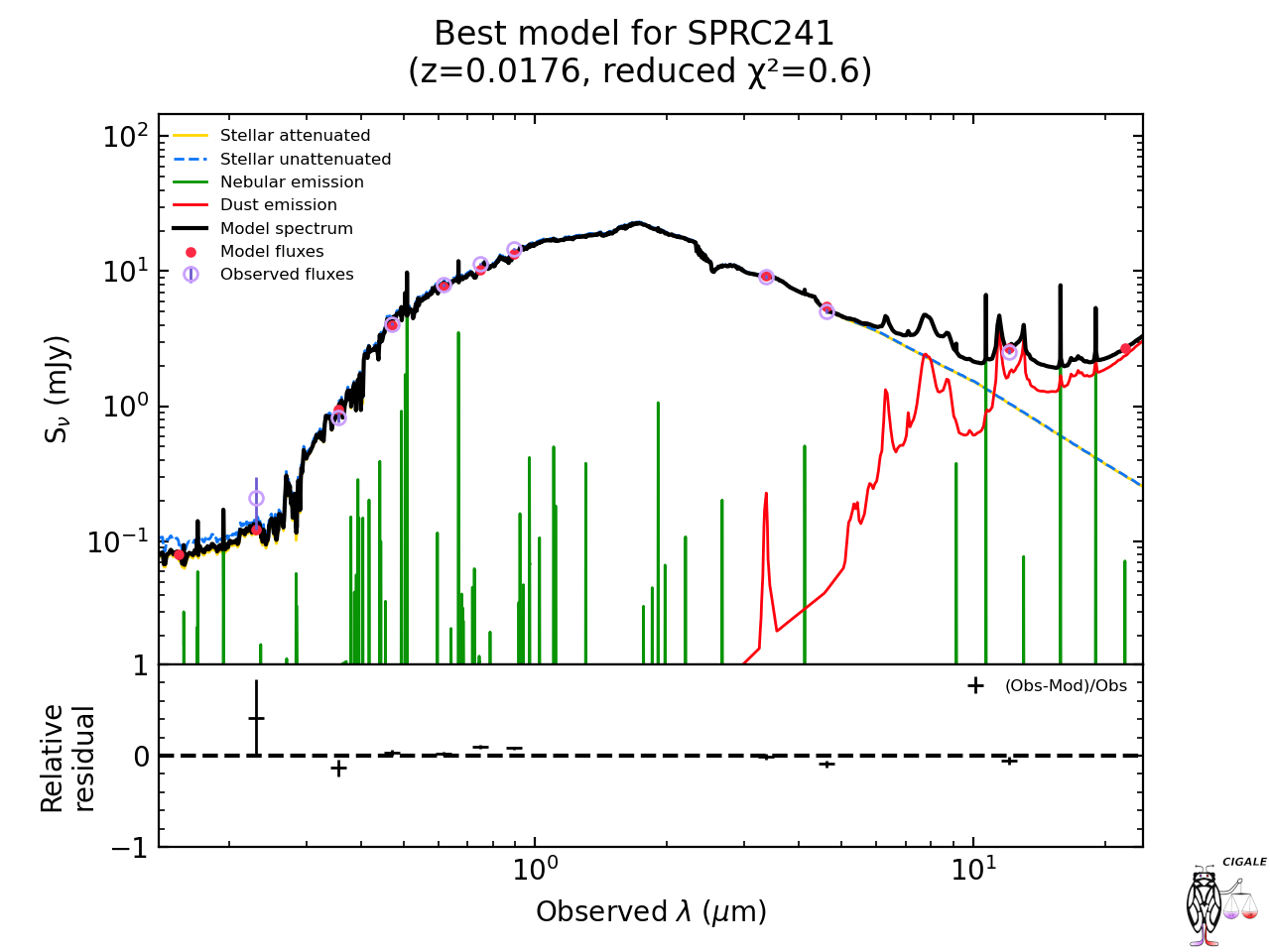}
    \includegraphics[width=0.45\linewidth]{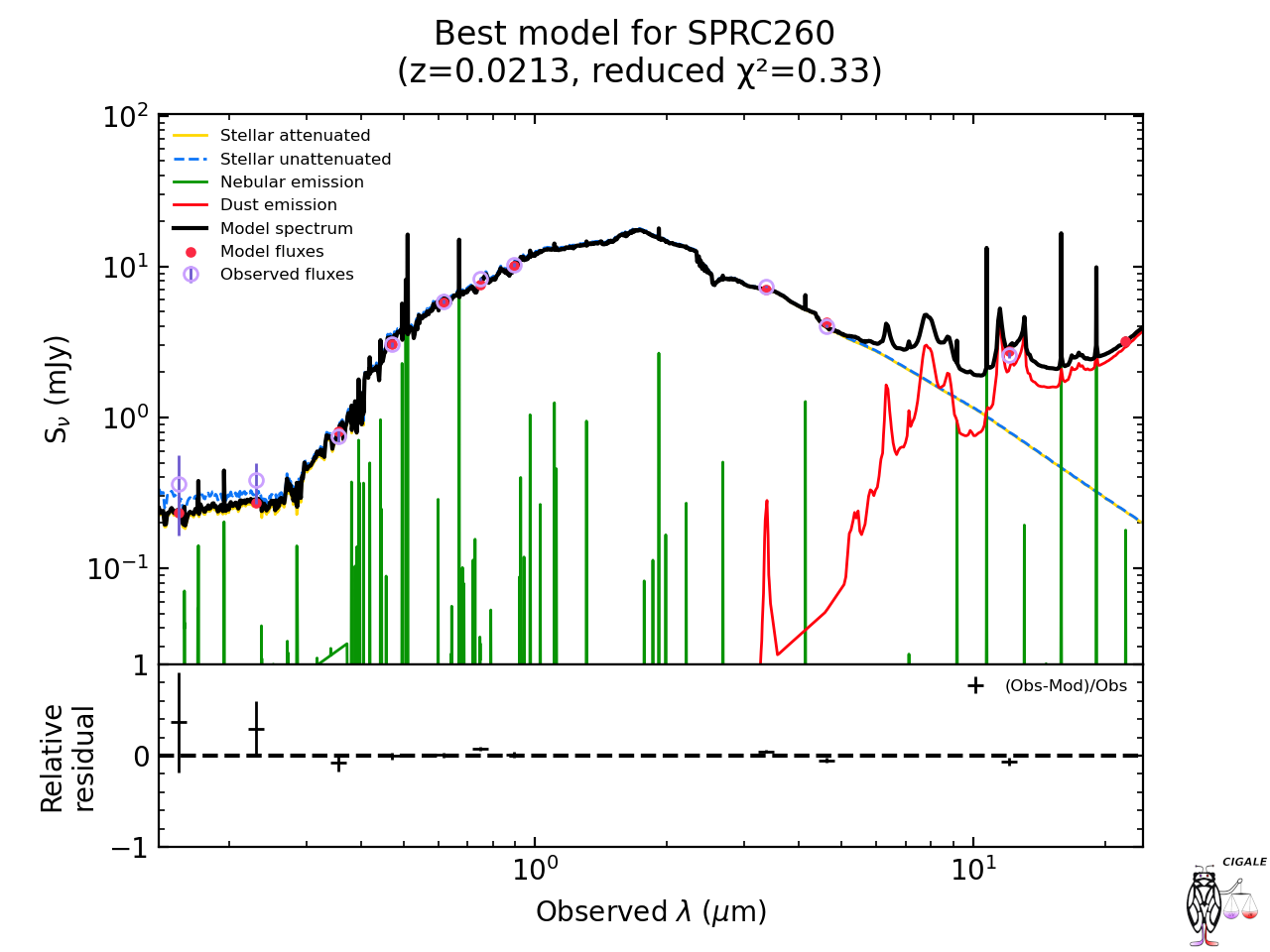}
    \caption{Best-fit SED models for two PRGs: SPRC241 and SPRC260. The SEDs for the other seven PRGs are in the Appendix \ref{seven_SED}., Fig \ref{fig:SED2}.}
    \label{fig:SED1}
\end{figure*}

\subsection{PRGs in the Cosmic Web: 3D Voronoi tessellation and DisPerSE}
\label{sec:cosmic_web}

The large-scale environments of the PRGs were characterized using two complementary approaches. Three-dimensional Voronoi tessellation, previously applied to
characterise the environments of SDSS galaxies at $z<0.1$
\citep{Dobrycheva2014, Elyiv2009, Vavilova2021b}, was used to quantify the relative
density of the surrounding galaxy distribution, while the DisPerSE
algorithm was applied to reconstruct the local filamentary skeleton and determine the position of each target relative to the nearest filament. The filament reconstruction tests whether the present-day position of each galaxy is geometrically compatible with a filamentary supply channel.
Interactive 3D visualizations of the Voronoi environments and the
fiducial DisPerSE reconstructions for the seven galaxies retained in
the quantitative analysis are available online.\footnote{
\url{https://olenakompaniiets.github.io/PRG-Cosmic-Web/}}

\subsubsection{Three-dimensional Voronoi tessellation}
\label{sec:voronoi_method}

We extended the environmental analysis previously applied to PRGs by \citet{Dobrycheva2025}, following the methodology described in detail by \citet{Kompaniiets2025_env}. For each target, neighboring galaxies within a radius of 30~Mpc were retained. This search scale was adopted to provide a sufficiently extended surrounding tracer distribution for the Voronoi cells of the targets included in the quantitative three-dimensional analysis to become bounded.

The three-dimensional Voronoi tessellation was constructed from the
comoving Cartesian coordinates of the target and its spectroscopic
neighbors. The local Voronoi density was defined as
$\Sigma_{\rm 3D,Vor}=V_{\rm Vor}^{-1}$, where $V_{\rm Vor}$ is the
volume of the bounded Voronoi cell associated with the target galaxy \citep{Elyiv2009, Dobrycheva2015, Vavilova2021, Melnyk2006}.
Following the density-based classification of
\citet{Kompaniiets2025_env}, we adopt
$\Sigma_{\rm 3D,Vor}<0.05~{\rm Mpc}^{-3}$ as the criterion for a
low-density environment. This density criterion complements the
filament-proximity and reconstruction-stability criteria used for
the final environmental classification. The Voronoi densities are
used primarily as relative environmental-density indicators within
the sample.

For IC1689 and SPRC241, with $z=0.0152$ and $z=0.0176$, respectively, the quantitative three-dimensional Voronoi density was not used. At these low redshifts, peculiar velocities constitute a comparatively larger fraction of the observed recession velocity and can produce substantial uncertainty in redshift-derived line-of-sight distances. We therefore adopted a conservative threshold of $z=0.02$ for the quantitative three-dimensional environmental analysis. IC1689 and SPRC241 were retained in the SED analysis but excluded from the quantitative three-dimensional environmental classification. The same low-redshift criterion was subsequently applied to the quantitative three-dimensional DisPerSE analysis.

\begin{figure}
    \centering
    \includegraphics[width=0.95\linewidth]{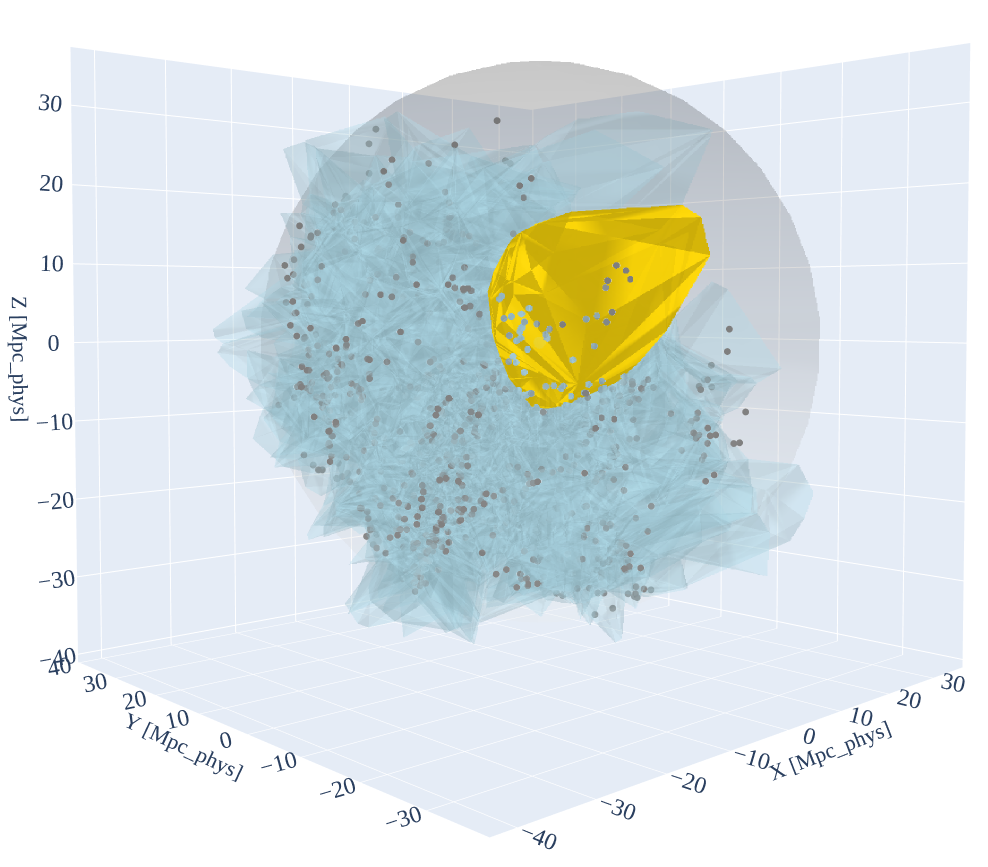}
    \caption{An example of the environment reconstruction based on the 3D Voronoi tesselation for SPRC179. }
    \label{fig:3D}
\end{figure}

\begin{table}
\caption{Three-dimensional Voronoi density estimates for the PRGs sample. Spectroscopic neighbors were searched within 30~Mpc around each target. For galaxies at $z<0.02$, the quantitative 3D Voronoi density was not used because redshift-space distances are increasingly affected by peculiar velocities at these low redshifts.}
\label{tab:voronoi_environment}

\centering
\begin{tabular}{lcc}
\hline
\hline

Name &
$z$ &
$\Sigma_{\rm 3D,Vor}$
\\

 &
 &
(Mpc$^{-3}$)
\\

\hline

IC1689
& 0.015200
& --
\\

SPRC027
& 0.048352
& 0.00350
\\

SPRC069
& 0.024652
& 0.01017
\\

SPRC056
& 0.055038
& 0.08969
\\

UGC7576
& 0.023407
& 0.01899
\\

SPRC260
& 0.021380
& 0.13087
\\

SPRC241
& 0.017600
& --
\\

UGC7388
& 0.021600
& 0.00985
\\

SPRC179
& 0.032100
& 0.000244
\\

\hline
\end{tabular}

\tablefoot{
$\Sigma_{\rm 3D,Vor}$ is the inverse-volume density derived from the
bounded three-dimensional Voronoi cell. A dash indicates that the
quantitative 3D Voronoi density was not used for the two targets with
$z<0.02$.
}
\end{table}

\subsubsection{Filament reconstruction with DisPerSE}
\label{sec:disperse_method}

To assess whether the present-day environment of each PRG is
geometrically compatible with filament-associated gas supply, we
reconstructed the local filamentary network using DisPerSE
\citep{Sousbie2011a,Sousbie2011b}. The input catalogues were drawn
from SDSS DR17 and restricted to spectroscopically classified
galaxies with reliable primary spectra
(\texttt{zWarning}=0 and \texttt{sciencePrimary}=1) and clean
photometric galaxy classifications
(\texttt{clean}=1 and \texttt{type}=3).
Repeated entries within $1''$ were removed.
To reduce contamination by the target itself and its deblended
components, we excluded sources within an angular radius equal
to the larger of $30''$ and three times the target's SDSS
$r$-band Petrosian radius containing 90\% of the flux, where
available. A minimum three-dimensional comoving separation of
0.05~cMpc from the target was also imposed.
The retained galaxy positions were converted to comoving
Cartesian coordinates, and the reconstruction volumes were
centered on the corresponding PRGs.
The Delaunay tessellation was constructed using the
\texttt{smooth} boundary option of DisPerSE.

Consistent with the Voronoi analysis, IC1689 and SPRC241 were excluded from the quantitative three-dimensional DisPerSE classification because their redshifts are below the adopted $z=0.02$ threshold. The robustness analysis described below therefore refers to the seven galaxies retained in the quantitative 3D Cosmic Web analysis.

We tested the sensitivity of the recovered filament geometry to both the size of the reconstructed volume and the persistence threshold. For each PRGs retained in the quantitative 3D analysis, DisPerSE was run within comoving radii of 30, 50, and 70~Mpc using persistence thresholds of $2\sigma$, $3\sigma$, $4\sigma$, and $5\sigma$. This produced a grid of 12 reconstruction configurations per galaxy. The $R=50$~Mpc, $3\sigma$ reconstruction was adopted as the fiducial configuration, while the complete grid was used to quantify the robustness of the inferred filament geometry.

For every reconstruction, we identified the nearest point on the DisPerSE filament skeleton and calculated the minimum three-dimensional comoving distance from the PRGs center to the corresponding filament polyline. This quantity is denoted by \(D_{\rm spine}\):
\begin{equation}
D_{\rm spine} = \min_{\boldsymbol{x}\in{\cal S}}
\left|
\boldsymbol{x}_{\rm PRGs}-\boldsymbol{x}
\right|
\end{equation}
where \({\cal S}\) denotes the reconstructed filament skeleton. Thus, \(D_{\rm spine}\) measures the distance to the filament spine itself and not the distance to the nearest neighboring galaxy. All distances used in the DisPerSE analysis are expressed in comoving coordinates.

The stability of $D_{\rm spine}$ was evaluated separately against
changes in persistence threshold and reconstruction volume.
The persistence test used the four thresholds at the fiducial
radius of 50~cMpc, while the volume test used the three radii
at the fiducial persistence threshold of $3\sigma$.
For each test, we defined the absolute and relative distance spreads as
\begin{equation}
\begin{aligned}
\Delta_D &=
P_{84}(D_{\rm spine})-P_{16}(D_{\rm spine}), \\
\delta_D &=
\frac{\Delta_D}
{\max\!\left(
|\operatorname{median}(D_{\rm spine})|,
0.1~{\rm cMpc}
\right)}.
\end{aligned}
\end{equation}
where $P_{16}$ and $P_{84}$ are the corresponding percentiles
of the successful reconstructions.
Both tests required
$\Delta_D\leq0.75$~cMpc and $\delta_D\leq0.35$.
At least three successful reconstructions were required for
the persistence test and at least two for the volume test.

For the volume test, we additionally calculated the maximum
three-dimensional displacement of the closest skeleton points
from their coordinate-wise median position.
Where evaluable, this displacement was required to be no larger
than 1~cMpc.
The fiducial reconstruction was further required to contain at
least 50 spectroscopic tracers, with tracers present in at least
six of the eight Cartesian octants centered on the target.
To identify solutions approaching the outer reconstruction
boundary, we required
$D_{\rm spine}^{\rm fid}/R_{\rm fid}\leq0.8$.
A system was assigned a robust filament-spine geometry only
when the fiducial distance was finite and all applicable
stability, sampling, and boundary requirements were satisfied.
These operational criteria quantify sensitivity to the tested
reconstruction choices.
For systems failing this criterion, we retained the full
minimum-to-maximum distance range across the 12 reconstructions
and discussed whether their qualitative filament proximity
remained unchanged.

Since DisPerSE reconstructs the topological skeleton of the density field and does not directly provide a physical filament boundary, we estimated an effective transverse radius, \(R_{\rm fil}\), from the distribution of neighboring galaxies around the corresponding filament spine. In cosmological simulations, \citet{GalarragaEspinosa2020} used radial galaxy-density profiles around DisPerSE spines to characterize the transverse structure of filaments.

For this purpose, the filament containing the closest skeleton point was first extended through adjacent DisPerSE arcs to construct a continuous local ridge. Individual DisPerSE arcs terminate at critical points, although the same ridge may continue through adjacent arcs. At a branching critical point, the continuation with the smallest change in local direction was selected, i.e. the adjacent arc whose outgoing tangent had the largest scalar product with the incoming filament tangent. The ridge was extended on both sides of the closest point until the local length required for the transverse-density measurement was reached or no connected continuation remained.

The neighboring SDSS galaxies were then expressed in local coordinates relative to this connected spine. For each galaxy, we determined its minimum transverse distance \(r_\perp\) from the spine and retained tracers whose closest point along the ridge fell within a finite longitudinal interval centered on the PRGs projection onto the spine. The fiducial configuration used a half-length of 5~comoving Mpc.

We constructed a cylindrical radial number-density profile around
the connected local filament ridge.
The longitudinal interval was centered on the target's projection
onto the ridge. When the available connected ridge was shorter
on either side, the interval was shortened symmetrically.
Profiles were evaluated only when the effective longitudinal
half-length was at least 2~cMpc.
For an annulus bounded by $r_i$ and $r_{i+1}$, the tracer density
and its counting uncertainty were estimated as
\begin{equation}
\rho_i=\frac{N_i}{V_i},
\qquad
\sigma_i=\frac{\sqrt{\max(N_i,1)}}{V_i},
\qquad
V_i=\pi(r_{i+1}^{2}-r_i^{2})L,
\end{equation}
where $N_i$ is the number of tracers in the annulus and $L$
is the effective longitudinal length.

The background was estimated from the outer radial bins using
an inverse-variance-weighted mean,
\begin{equation}
\rho_{\rm bg}
=
\frac{\sum_{i\in{\cal B}}w_i\rho_i}
{\sum_{i\in{\cal B}}w_i},
\qquad
\sigma_{\rm bg}
=
\left(\sum_{i\in{\cal B}}w_i\right)^{-1/2},
\qquad
w_i=\sigma_i^{-2},
\end{equation}
where ${\cal B}$ denotes bins whose centers lie within the
adopted background interval.
A profile required at least 20 local tracers within the
maximum transverse radius and at least eight tracers in the
background bins, with at least two background bins available.

The effective filament radius, $R_{\rm fil}$, was defined as
the inner edge of the first pair of consecutive radial bins
satisfying
\begin{equation}
\rho_i\leq
\rho_{\rm bg}+\sqrt{\sigma_i^2+\sigma_{\rm bg}^2}.
\end{equation}
The search started outside the innermost radial bin, and the
first bin of the accepted pair was required to have its center
inside the inner boundary of the background interval.
This operational criterion identifies where the profile no
longer shows a significant excess above the estimated local
background.
The resulting $R_{\rm fil}$ characterizes the transverse extent
of the galaxy-traced filament environment.

Three profile configurations were evaluated for every DisPerSE
reconstruction:
\begin{align}
(L/2,r_{\max},\Delta r,r_{\rm bg})
&=(5,6,0.30,3.5\!-\!6), \nonumber\\
&=(5,5,0.25,3\!-\!5), \nonumber\\
&=(7,6,0.40,3.5\!-\!6),
\end{align}
with all lengths expressed in comoving Mpc.
Here $L/2$ is the requested longitudinal half-length,
$r_{\max}$ is the maximum transverse radius,
$\Delta r$ is the radial-bin width, and $r_{\rm bg}$
is the background interval.
Consequently, up to 36 individual width estimates were
available per target.

For each finite width estimate, we generated 300 bootstrap
realizations by resampling the local tracer population with
replacement and recomputing the profile.
The lower and upper width bounds were defined by the 16th
and 84th percentiles of the finite bootstrap estimates.
An interval was accepted when at least 60 of the 300
realizations yielded a finite width.
All finite width estimates contributing to the present
classification satisfied this bootstrap requirement.
Configurations with insufficient ridge length, too few
local or background tracers, or no qualifying pair of radial
bins were retained as unconstrained measurements.
No finite filament radius was assigned to those configurations.

The distance between the target galaxy and the effective filament edge was defined as
\begin{equation}
D_{\rm edge}=D_{\rm spine}-R_{\rm fil}.
\end{equation}
Negative values of \(D_{\rm edge}\) indicate that the galaxy lies within the estimated transverse extent of the reconstructed filament, while positive values indicate that it lies outside this region.

To account for the spatial extent of the host dark-matter halo, we introduced an additional geometrical criterion based on its virial radius. Halo masses were inferred from the CIGALE stellar masses using the redshift-dependent stellar-to-halo mass relation of \citet{Moster2013}. The corresponding virial radius was calculated as
\begin{equation}
R_{200} = \left( \frac{3M_{200}} {4\pi\,200\,\rho_{\rm crit}(z)} \right)^{1/3},
\end{equation}
where \(M_{200}\) is the inferred halo mass and \(\rho_{\rm crit}(z)\) is the critical density of the Universe at the galaxy redshift.

Because $R_{200}$ obtained from the above expression is a physical
radius, it was converted to the corresponding comoving scale,
$R_{200,\rm com}=R_{200}(1+z)$, before comparison with the
DisPerSE distances.
Uncertainties in the CIGALE stellar masses were propagated
through the stellar-to-halo mass conversion using 500 Monte
Carlo realizations per galaxy.
Stellar masses were sampled from a Gaussian distribution
with the reported mean and standard deviation, with a lower
bound of $10^7~M_\odot$ imposed on the sampled values.
Each sampled stellar mass was converted to a halo mass using
the redshift-dependent relation of \citet{Moster2013}.
As an additional sensitivity test, a Gaussian perturbation
with a standard deviation of 0.15~dex was applied to
$\log_{10}M_{\rm halo}$.
This adopted halo-mass perturbation is distinct from the
intrinsic scatter in stellar mass at fixed halo mass in
\citet{Moster2013}.
The resulting 16th and 84th percentiles of
$R_{200,\rm com}$ were used as its lower and upper bounds
in the geometrical tests.
These intervals describe the uncertainty contributions
included in this analysis.

A galaxy was classified as geometrically compatible with present-day filament-fed gas supply when
\begin{equation}
D_{\rm spine}\leq R_{\rm fil}+2R_{200,\rm com},
\end{equation}
or, equivalently,
\begin{equation}
D_{\rm edge}\leq 2R_{200,\rm com}.
\end{equation}
The spine position is determined from the spectroscopic tracers, whereas the final association criterion also depends on the stellar mass inferred from the integrated SED through the estimated halo radius. The adopted threshold is an empirical geometrical filter introduced in this work. It identifies systems whose host halos overlap with, or closely approach, the effective transverse region associated with the nearest reconstructed filament. It should not be interpreted as a sharply defined physical boundary for gas accretion.

The criterion was evaluated over the full reconstruction and filament-width ensemble instead of being based solely on the fiducial \(3\sigma\) solution. For each finite \(R_{\rm fil}\) estimate we calculated the corresponding value of \(D_{\rm spine}-(R_{\rm fil}+2R_{200,\rm com})\). We further constructed a conservative uncertainty envelope using the lower bootstrap bound of \(R_{\rm fil}\) together with the lower \(R_{200,\rm com}\) estimate for the most restrictive filament-association test, and the upper bounds of both quantities for the most permissive test. An individual configuration was therefore considered robustly filament-associated when
\begin{equation}
D_{\rm spine}\leq R_{\rm fil,low}+2R_{200,\rm com}^{\rm low}
\end{equation}
whereas it was considered robustly filament-distant when
\begin{equation}
D_{\rm spine}>R_{\rm fil,high}+2R_{200,\rm com}^{\rm high}.
\end{equation}

Configurations for which the uncertainty interval crossed the geometrical boundary were treated as ambiguous. The classification remains unchanged when the uncertainty in \(R_{\rm fil}\) does not cross the geometrical boundary.

The final classification combined the filament-spine stability
test with the individual uncertainty-envelope results.
The fractions $f_{\rm assoc}$ and $f_{\rm distant}$ were
calculated relative to the number of evaluable individual
width estimates, $N_{\rm eval}$.
A system was classified as filament-associated when its spine
geometry was robust and at least 75\% of the evaluable estimates
remained associated under the restrictive envelope test.
A system was classified as filament-distant in a low-density
environment when its spine geometry was robust, at least 75\%
of the evaluable estimates remained distant under the permissive
envelope test, and
$\Sigma_{\rm 3D,Vor}<0.05~{\rm Mpc}^{-3}$.
Systems with robust spine geometry but no evaluable width
estimate, or with intermediate envelope fractions, were assigned
to the transition class.
Failure of the spine-stability criterion took precedence over
the width-based result and led to the environment-uncertain class.

\begin{figure}
    \centering
    \includegraphics[width=1.0\linewidth]{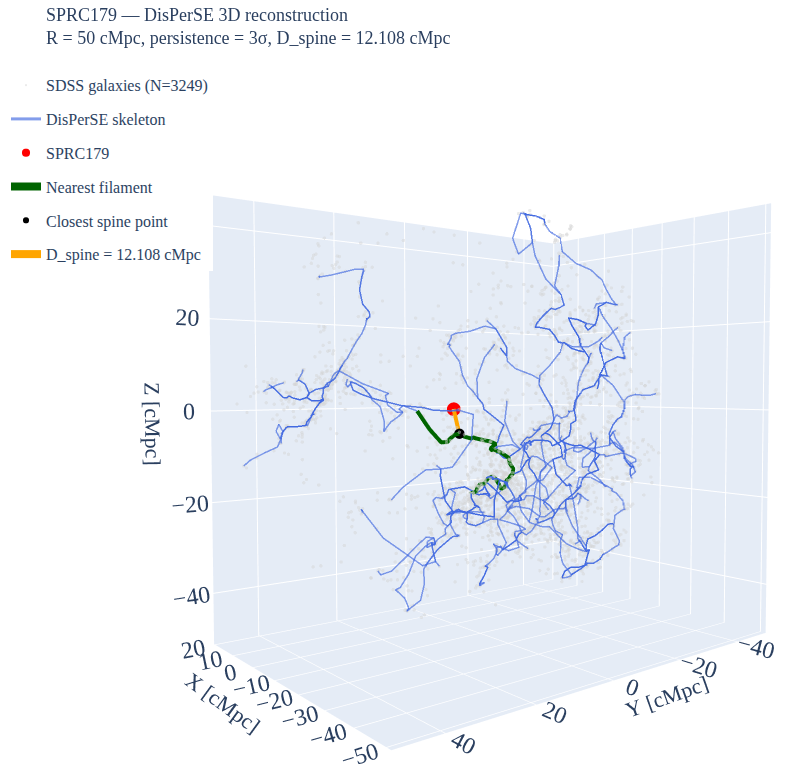}
    \caption{An example of DisPerSE reconstruction for SPRC179.  }
    \label{fig:DisPerSE}
\end{figure}

The 75\% threshold is an operational measure of classification
stability across the reconstruction and profile choices.
The 36 attempted configurations reuse the same underlying
tracer data and are correlated; their fractions are therefore
reported as sensitivity diagnostics.
The number of evaluable configurations is given explicitly
to distinguish agreement among available measurements from
the overall success rate of the width estimation.
The results are presented in Table~\ref{tab:disperse_results},
and an example reconstruction is shown in
Fig.~\ref{fig:DisPerSE}.

For targets without a formally robust \(D_{\rm spine}\), we additionally report the range of spine distances obtained across the reconstruction grid. Failure of the formal stability criterion can arise from a change in the identity or location of the nearest reconstructed filament or from a smaller displacement of the spine that leaves the qualitative proximity of the galaxy to the Cosmic Web unchanged. Such cases are discussed individually.

The reconstructed coordinates are derived from spectroscopic
redshifts.
The analysis does not apply a peculiar-velocity or Finger-of-God
correction, and the octant and radial-boundary diagnostics do
not constitute a full survey-completeness correction.
The robustness classification therefore refers to the tested
reconstruction and profile choices in redshift space.

Satisfaction of the adopted geometrical criterion indicates
consistency between the present-day reconstructed environment
and a possible filament-associated supply pathway.
For systems classified as filament-distant, the reconstructed
geometry provides no comparable present-day spatial support.
An environment-uncertain classification indicates that the
available reconstruction does not support a stable geometrical
assignment.
These classifications neither detect gas inflow nor reconstruct
the environment at the epoch of polar-ring formation.
They are combined with the integrated stellar-population
constraints when discussing the possible formation pathways
of the individual PRGs.

\begin{table*}
\centering
\small
\setlength{\tabcolsep}{4pt}
\renewcommand{\arraystretch}{1.15}
\caption{Galaxy properties and results of the local DisPerSE filament analysis.}
\label{tab:disperse_results}

\begin{tabular*}{\textwidth}{@{\extracolsep{\fill}}lccccccc@{}}
\hline\hline
\multicolumn{8}{@{}l}{\textit{Galaxy properties and spectroscopic tracers}} \\
\hline
Name &
RA &
Dec &
$z$ &
$M_\star$ &
$M_{\rm halo}$ &
$R_{200,\rm com}$ &
$N_{\rm nei}^{\rm fid}$ \\
&
(deg) &
(deg) &
&
($M_\odot$) &
($M_\odot$) &
(cMpc) &
\\
\hline
IC1689
& 20.9494 & 33.0553 & 0.015200
& $1.67\times10^{10}$ & $4.61\times10^{11}$ & 0.165 & -- \\
SPRC027
& 176.1834 & 23.1625 & 0.048352
& $1.98\times10^{10}$ & $5.53\times10^{11}$ & 0.179 & 3004 \\
SPRC069
& 312.0236 & 0.0688 & 0.024652
& $1.46\times10^{10}$ & $4.21\times10^{11}$ & 0.161 & 620 \\
SPRC056
& 227.8087 & 37.0438 & 0.055038
& $8.95\times10^{10}$ & $8.10\times10^{12}$ & 0.439 & 2267 \\
UGC7576
& 186.9244 & 28.6981 & 0.023407
& $2.14\times10^{10}$ & $5.78\times10^{11}$ & 0.179 & 8496 \\
SPRC260
& 176.3761 & 9.7291 & 0.021380
& $1.75\times10^{10}$ & $4.83\times10^{11}$ & 0.168 & 6737 \\
SPRC241
& 40.4875 & $-6.7923$ & 0.017600
& $1.60\times10^{10}$ & $4.48\times10^{11}$ & 0.163 & -- \\
UGC7388
& 185.0656	& 33.6610 & 0.021600
& $3.31\times10^{10}$ & $9.57\times10^{11}$ & 0.211 & 8273 \\
SPRC179
& 255.7866 & 46.3736 & 0.032100
& $6.76\times10^{10}$ & $3.98\times10^{12}$ & 0.342 & 3249 \\
\hline
\end{tabular*}

\par\vspace{0.8em}

\begin{tabular*}{\textwidth}{@{\extracolsep{\fill}}lccccccl@{}}
\hline\hline
\multicolumn{8}{@{}l}{\textit{Filament geometry and environmental classification}} \\
\hline
Name &
$D_{\rm spine}^{\rm fid}$ &
$D_{\rm spine}^{\rm range}$ &
$R_{\rm fil}^{\rm range}$ &
$N_{\rm eval}$ &
$f_{\rm assoc}$ &
$f_{\rm distant}$ &
Environmental class \\
&
(cMpc) &
(cMpc) &
(cMpc) &
&
&
&
\\
\hline
IC1689
& -- & -- & -- & -- & -- & --
& Excluded from 3D analysis \\
SPRC027
& 6.915 & 6.915--6.915 & 0.25--2.80 & 13/36 & 0.00 & 1.00
& \shortstack[l]{Filament-distant,\\low-density} \\
SPRC069
& 0.985 & 0.296--1.204 & -- & 0/36 & -- & --
& Environment-uncertain \\
SPRC056
& 0.337 & 0.337--0.337 & 2.80--3.00 & 4/36 & 1.00 & 0.00
& Filament-associated \\
UGC7576
& 3.632 & 3.632--3.632 & 0.80--2.80 & 36/36 & 0.00 & 1.00
& \shortstack[l]{Filament-distant,\\low-density} \\
SPRC260
& 0.993 & 0.993--0.993 & 0.90--2.80 & 36/36 & 0.667 & 0.00
& Transition \\
SPRC241
& -- & -- & -- & -- & -- & --
& Excluded from 3D analysis \\
UGC7388
& 5.368 & 5.368--5.368 & 0.90--3.20 & 36/36 & 0.00 & 1.00
& \shortstack[l]{Filament-distant,\\low-density} \\
SPRC179
& 12.108 & 10.137--13.999 & 0.25--3.20 & 30/36 & 0.00 & 1.00
& Environment-uncertain$^{a}$ \\
\hline
\end{tabular*}

\tablefoot{
Stellar masses are from CIGALE. Nominal halo masses are inferred
from the stellar-to-halo mass relation, and $R_{200,\rm com}$ is
the comoving virial radius. The superscript \textit{fid} refers
to the $R=50$~cMpc, $3\sigma$ reconstruction.
$N_{\rm nei}^{\rm fid}$ is its spectroscopic tracer count, and
$D_{\rm spine}^{\rm fid}$ is the distance to its nearest filament
spine. The ranges span the 12 reconstructions for $D_{\rm spine}$
and the finite width estimates for $R_{\rm fil}$.
$N_{\rm eval}$ counts evaluable estimates out of 36 attempts
(three profiles per reconstruction).
The fractions $f_{\rm assoc}$ and $f_{\rm distant}$ summarize
agreement among these correlated estimates under the restrictive
and permissive envelope tests, respectively. The envelope uses
the 16th and 84th percentiles of the bootstrap filament widths
and Monte Carlo halo radii.

Low density denotes $\Sigma_{\rm 3D,Vor}<0.05~{\rm Mpc}^{-3}$.
Environment-uncertain denotes failure of the spine-stability
criterion. Dashes indicate unavailable or unevaluated quantities.
Targets at $z<0.02$ were excluded because their reconstructions
are more sensitive to peculiar velocities and the observer boundary.

$^{a}$ SPRC179 fails the spine-stability criterion but remains
far from the nearest reconstructed filament in all 12 runs.
All 30 evaluable profiles satisfy the filament-distance criterion.
}
\end{table*}

\section{Results and discussion} \label{discuss}

\subsection{What does the integrated photometry constrain?}

The inferred present-day SFRs range from $0.014\pm0.012~M_\odot\,{\rm yr}^{-1}$ for IC1689 to $0.944\pm0.289~M_\odot\,{\rm yr}^{-1}$ for UGC7388. In the mock analysis, the input and recovered values correlate more closely for SFR and stellar mass ($r^2=0.96$ and 0.98) than for the secondary-component decline timescale ($r^2=0.32$). Mock test assesses parameter recovery within the adopted
two-component SFH family, since the same model grid is used to
generate and fit the mock photometry. Comparisons across
parametric SFH families show that broad-band photometry can
admit different star-formation histories \citep{Carnall2019}.

Within the adopted \texttt{sfh2exp} parameterization, the second
exponentially declining component represents a later episode
of stellar mass growth.
Its characteristic timescale, $\tau_2$, ranges in Bayesian
central estimate from approximately $1.23$ to $1.86$~Gyr.
UGC7388 has the largest value,
$\tau_2=1.86\pm1.41$~Gyr, while SPRC179 has the smallest,
$\tau_2=1.23\pm1.21$~Gyr; the remaining galaxies have
central estimates between approximately $1.34$ and $1.74$~Gyr.
However, the uncertainties are large,
approximately $1.21$--$1.41$~Gyr, and the posterior estimates
overlap strongly across the sample.
The overlapping uncertainties preclude a well-resolved
evolutionary ordering of the galaxies by $\tau_2$.

The interpretation of these timescales is further limited by the
available wavelength coverage and by the intrinsic degeneracy of
recovering detailed SFHs from integrated broad-band photometry. The SEDs
combine GALEX, SDSS, and WISE measurements but contain no far-infrared
observations sampling the thermal dust-emission peak. Consequently, the
total dust-reprocessed luminosity and the obscured contribution to the
current star formation are less directly constrained. In addition,
changes in the age and mass fraction of the secondary component, dust
attenuation, and its exponential timescale can produce similar UV-to-IR
SEDs. The weak recovery of $\tau_2$ in the CIGALE mock analysis
(Appendix~\ref{app:mock}) independently confirms this limitation. Within
the adopted model, the $\tau_2$ estimates characterize the secondary SFH
component on broad timescales, with limited precision in the duration
of individual star-formation episodes. Since the photometry is integrated over the entire galaxy,
neither $\tau_2$ nor the inferred SFR can be identified
uniquely with the polar ring itself.

\subsection{Which filament associations are robust?} For the seven galaxies retained in the quantitative
three-dimensional analysis, the Voronoi densities span
$\Sigma_{\rm 3D,Vor}=2.44\times10^{-4}$--$1.31\times10^{-1}~
{\rm Mpc}^{-3}$, demonstrating substantial variation in the local galaxy
density across the sample. These values characterize relative
environmental densities within the sample. IC1689 and SPRC241 were
retained in the SED analysis but excluded from the quantitative
three-dimensional environmental classification because their low
redshifts make redshift-derived radial positions more sensitive to
peculiar velocities and place the required reconstruction volumes close
to the observer boundary.

For the remaining seven galaxies, the DisPerSE reconstruction
reveals a wide range of geometries relative to the filamentary
skeleton.
SPRC056 provides the clearest filament-associated configuration.
Its nearest-spine distance remains
$D_{\rm spine}=0.337$~cMpc in all 12 reconstructions.
Only four of the 36 attempted transverse profiles yield finite
widths, $R_{\rm fil}=2.8$--$3.0$~cMpc; the other 32 fail the
minimum-background-tracer requirement.
All four finite estimates remain associated under the restrictive
uncertainty-envelope test.

The geometrical association criterion for SPRC056 is also
satisfied when the effective filament radius is set to zero:
$D_{\rm spine}=0.337$~cMpc is smaller than twice the lower
halo-radius estimate,
$2R_{200,\rm com}^{16}=0.730$~cMpc.
Within the adopted geometrical criterion and halo-mass
sensitivity test, its association therefore does not depend
on the poorly constrained filament width.
The nearby spine is stable across the reconstruction grid,
while the transverse extent of its galaxy-traced environment
remains uncertain.

SPRC027 and UGC7576 represent the clearest filament-distant cases. Their
spine distances remain stable at $6.915$ and $3.632$~cMpc, respectively,
over the complete reconstruction grid. All 13 evaluable transverse
profiles for SPRC027 and all 36 for UGC7576 remain outside the most
permissive filament-plus-halo envelope, and their Voronoi densities are
among the lower values measured in the sample. The present-day geometry
therefore provides no direct spatial support for an association with the
nearest prominent reconstructed filament.

For UGC7576, spatially resolved $B$, $V$, and $R_{\rm C}$
surface photometry provides a basis for comparing the colours
of the host and polar ring \citep{Reshetnikov1994}.
The subsolar gas-phase metallicity of the ring and the absence
of a significant abundance gradient are consistent with the
acquisition of external material \citep{Spavone2011}.
Our reconstruction places UGC7576 well outside the estimated
extent of the nearest prominent filament at the observed epoch.
The origin of the polar material therefore remains open, with
earlier filamentary accretion, tidal transfer, and acquisition
from a gas-rich satellite all remaining viable.

For SPRC260, stellar kinematics in the host and the ionized-gas
velocity field of the ring indicate a mutual inclination of
approximately $89^\circ$ \citep{Khoperskov2013}.
The nearest filament spine lies at 0.993~cMpc in all 12
reconstructions. Of the 36 evaluable profile configurations,
24 satisfy the restrictive association criterion and the remaining
12 are ambiguous. The transition classification thus refers to
the proximity of this kinematically confirmed polar-ring galaxy
to the effective filament boundary.

The remaining systems demonstrate the importance of testing the
robustness of the filament reconstruction. SPRC069 has a fiducial
$D_{\rm spine}=0.985$~cMpc, but the full reconstruction grid spans
$0.296$--$1.204$~cMpc, and no finite transverse width is recovered.
Its environmental interpretation is therefore sensitive to the adopted
DisPerSE configuration, and the system is classified as
environment-uncertain.

UGC7388 lies 5.368~cMpc from the nearest filament spine in all
12 reconstructions. All 36 evaluable transverse profiles place
it outside the most permissive filament-plus-halo envelope,
and its Voronoi density is low. Deep Gemini imaging and
spectroscopy revealed two faint, extended loops around the galaxy
and detected large-scale rotation in the brighter loop
\citep{FaundezAbans2009}. Those authors proposed that the loops
could be debris from the tidal disruption of a companion.
The loop morphology thus provides independent evidence for
considering a tidal origin. Our reconstruction adds that UGC7388
is distant from the nearest prominent filament today, while its
environment at the epoch of loop formation remains unknown.

SPRC179 formally also fails the adopted spine-stability requirement, but
its qualitative geometry is different from that of SPRC069.
Across all 12 reconstructions, it remains far from the filamentary
skeleton, with $D_{\rm spine}=10.137$--$13.999$~cMpc. All 30 evaluable
transverse-width configurations are classified as robustly
filament-distant. Its exact nearest-spine distance is therefore
reconstruction-dependent, while the conclusion that the galaxy lies far
from the reconstructed prominent filament network is stable.

\subsection{What do the two measurements imply for polar-ring formation?}

SPRC056 and UGC7576 illustrate the limits of the combined interpretation. Their nearest-spine distances remain stable across all 12 reconstructions, at 0.337 and 3.632~cMpc, respectively. Yet their integrated SEDs yield nearly identical estimates of the secondary-component decline timescale, $1.56\pm1.33$ and $1.55\pm1.34$~Gyr. The environmental analysis distinguishes their present-day geometries, while the photometry permits a similarly broad range of decline timescales in both systems. The available data therefore leave the relation between $\tau_2$ and filament proximity unconstrained.

Our interpretation is broadly consistent with the H\,I study of
\citet{Yu2026}, which favors external gas acquisition in polar-ring
galaxies. For SPRC260, the asymmetric H\,I profile and the detection
of a gas-bearing companion motivate consideration of an
interaction-related origin alongside filamentary supply. However,
these observations do not identify the source of the polar material.
Similarly, the H\,I reservoirs in the filament-distant systems
SPRC027 and UGC7576 remain compatible with external acquisition
without requiring an association with the nearest filament at the
observed epoch.

More precise UV photometry would improve constraints on young stellar populations, while far-infrared measurements of the dust-emission peak would constrain obscured star formation \citep{Conroy2013,Boquien2019}. Separate photometry and spatially resolved stellar spectroscopy of the ring and host would test whether the later episode in the integrated SFH can be associated with the polar ring
\citep{Coccato2014,Iodice2015}. For systems with established
polar kinematics, such as SPRC260, gas-phase abundance
measurements across the ring and in nearby gas-rich companions
would help assess a possible donor. Resolved H\,I maps could
provide a complementary test by tracing gaseous bridges or
tidal debris. A larger PRG sample and control galaxies matched in mass and redshift, analyzed with the same reconstruction procedure, would establish whether the observed distribution of filament distances is characteristic of PRGs.

\section{Conclusion} \label{conclusion}

We assessed the constraints on polar-ring formation from integrated UV-to-IR photometry of nine CIPRG galaxies and from three-dimensional environmental reconstructions for seven of them.

The inferred present-day SFRs span
$0.014$--$0.944~M_\odot\,{\rm yr}^{-1}$,
while the Bayesian central estimates of the
secondary-component e-folding timescale span approximately
$1.23$--$1.86$~Gyr.
The uncertainties on $\tau_2$ are, however, approximately
$1.21$--$1.41$~Gyr, and the CIGALE mock analysis confirms
that this parameter is only weakly recovered from the
available broad-band photometry.
Within the adopted two-component SFH model, $\tau_2$
characterizes the exponential decline of the secondary
star-formation component, and its uncertainty limits
discrimination among possible formation scenarios.

IC1689 and SPRC241 were retained in the SED analysis but excluded from
the quantitative three-dimensional environmental classification. Among
the remaining seven systems, SPRC056 has a robustly reconstructed nearby
filament spine and is classified as filament-associated, although its
transverse filament width is poorly constrained. SPRC260 occupies a
transition regime; SPRC027, UGC7576, and UGC7388 are robustly filament-distant;
SPRC069 has a reconstruction-sensitive nearest-filament
geometry; and SPRC179 remains consistently far from the reconstructed
filament network despite failing the formal spine-stability criterion.

For this sample, the combined analysis provides a spatial constraint on one proposed formation pathway. It identifies galaxies that lie close to a prominent filament today and systems that remain distant across the tested reconstructions. The available integrated photometry does not resolve differences in the decline timescale of the secondary star-formation component, leaving its relation to filament proximity unconstrained. For UGC7388, the reconstructed distance to the nearest filament complements the tidal interpretation suggested by its stellar
loops. For SPRC260, the nearly perpendicular rotation of the
host and ring is well established, while both the source of the
polar gas and its connection to the nearby filament remain open.

\begin{acknowledgements}
This study was supported by the research project of young scientists
of the National Academy of Sciences of Ukraine for 2025--2026
(registration number 0125U002943).
O. Kompaniiets and O. Hetmantsev acknowledge scholarship support
from the National Academy of Sciences of Ukraine.
I. Izviekova acknowledges support through the Scholarship of the
President of Ukraine. This research made use of publicly available data from the GALEX,
SDSS, and NEOWISE archives, as well as the NASA Astrophysics
Data System (ADS).
\end{acknowledgements}

\bibliographystyle{aa} 
\bibliography{library} 

\onecolumn
\appendix
\nolinenumbers

\section{Best-fit spectral energy distribution for the PRGs sample} \label{seven_SED}
\begin{figure}[!htbp]
\centering
\begin{tabular}{@{}cc@{}}
 \includegraphics[width=0.39\textwidth]{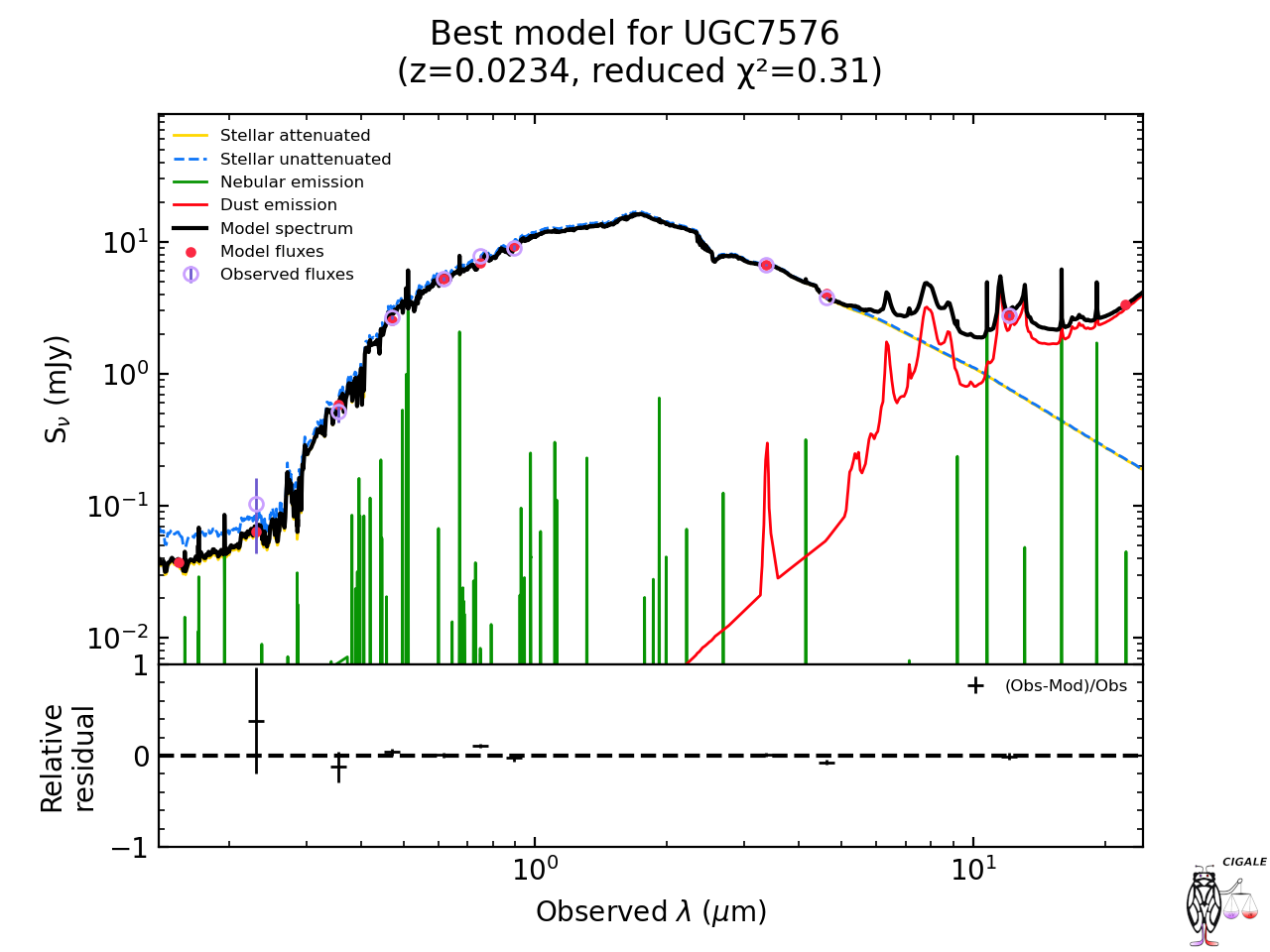}    & \includegraphics[width=0.39\textwidth]{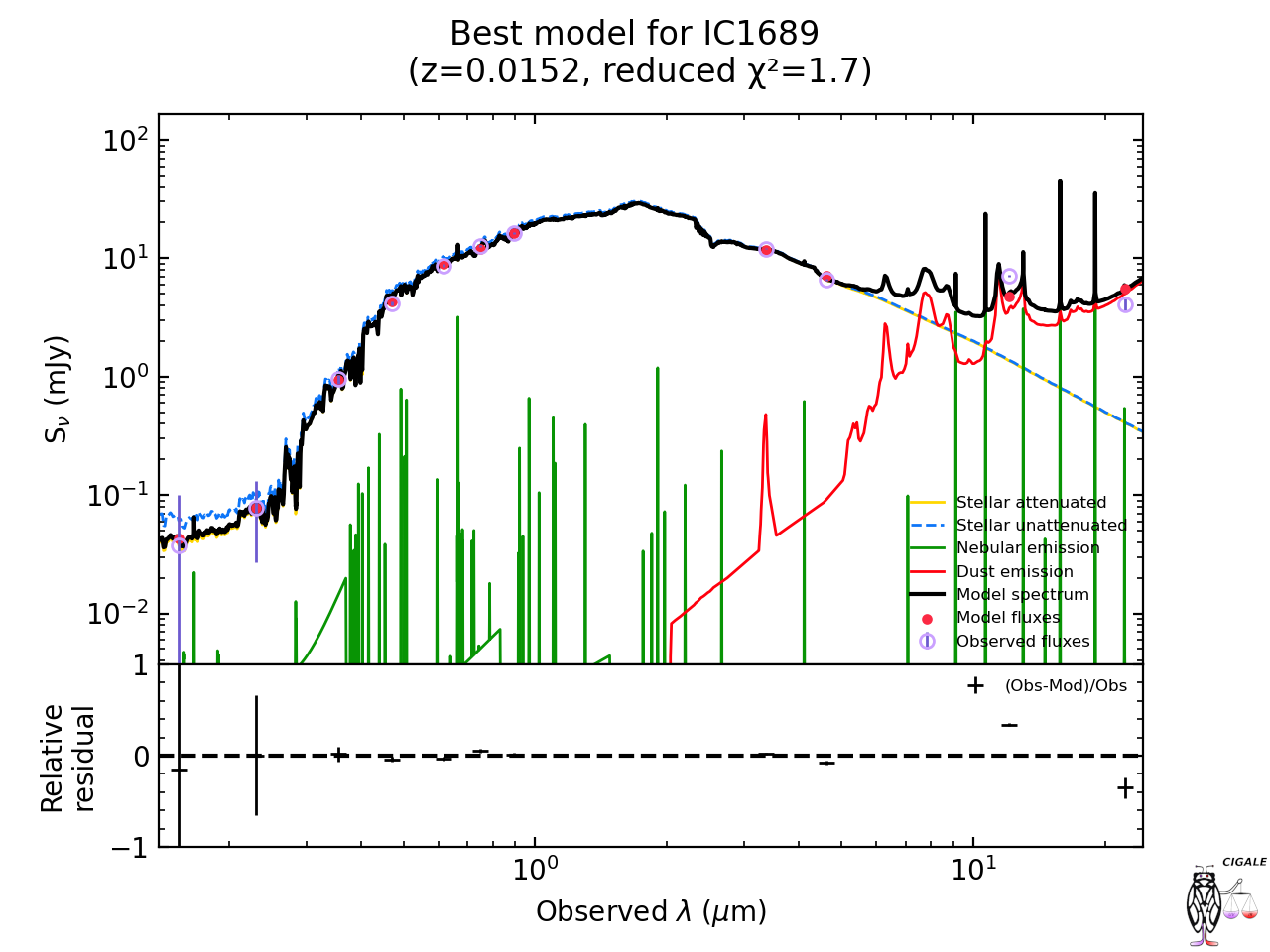}  \\
   \includegraphics[width=0.39\textwidth]{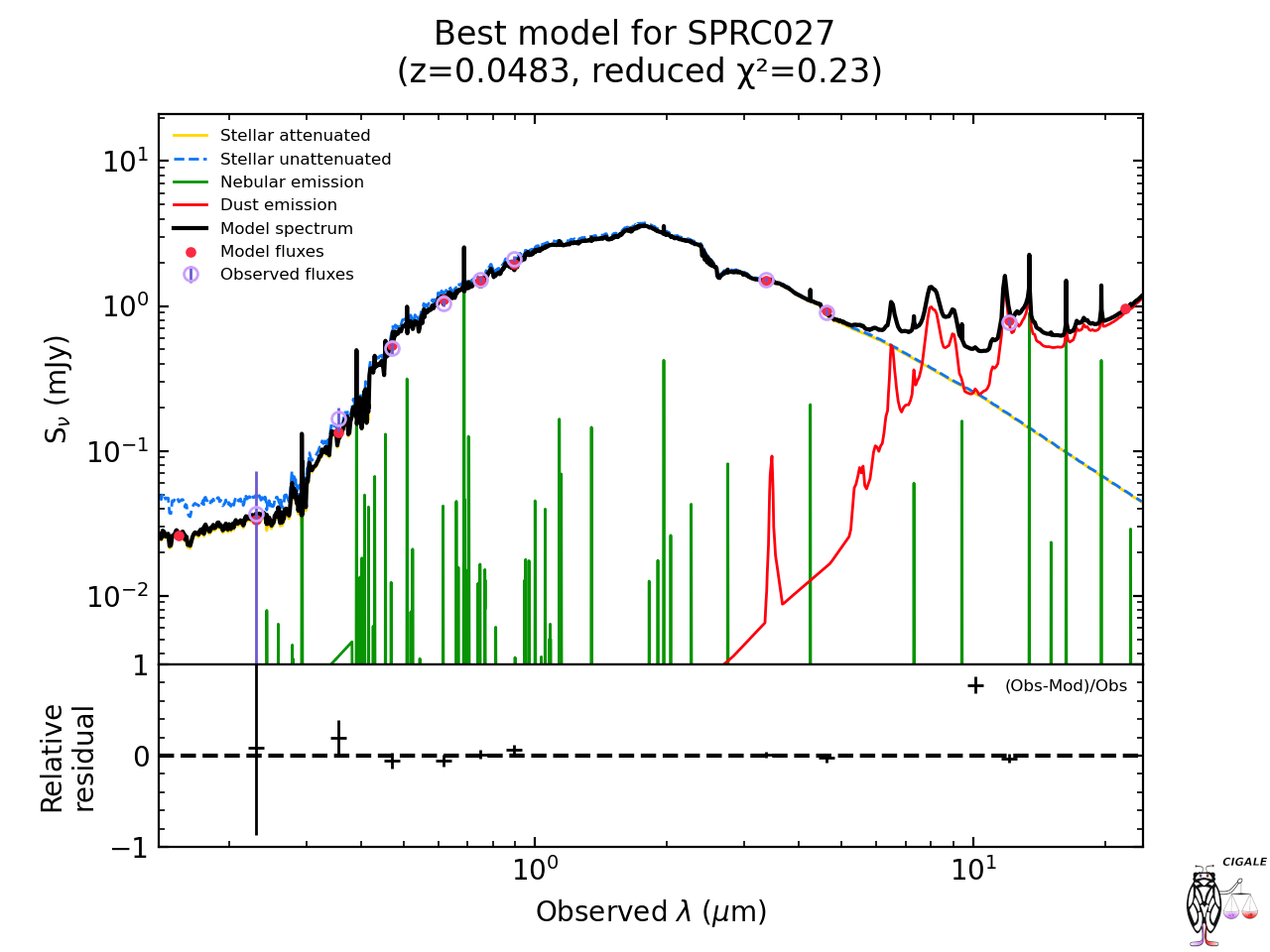}   &  \includegraphics[width=0.39\textwidth]{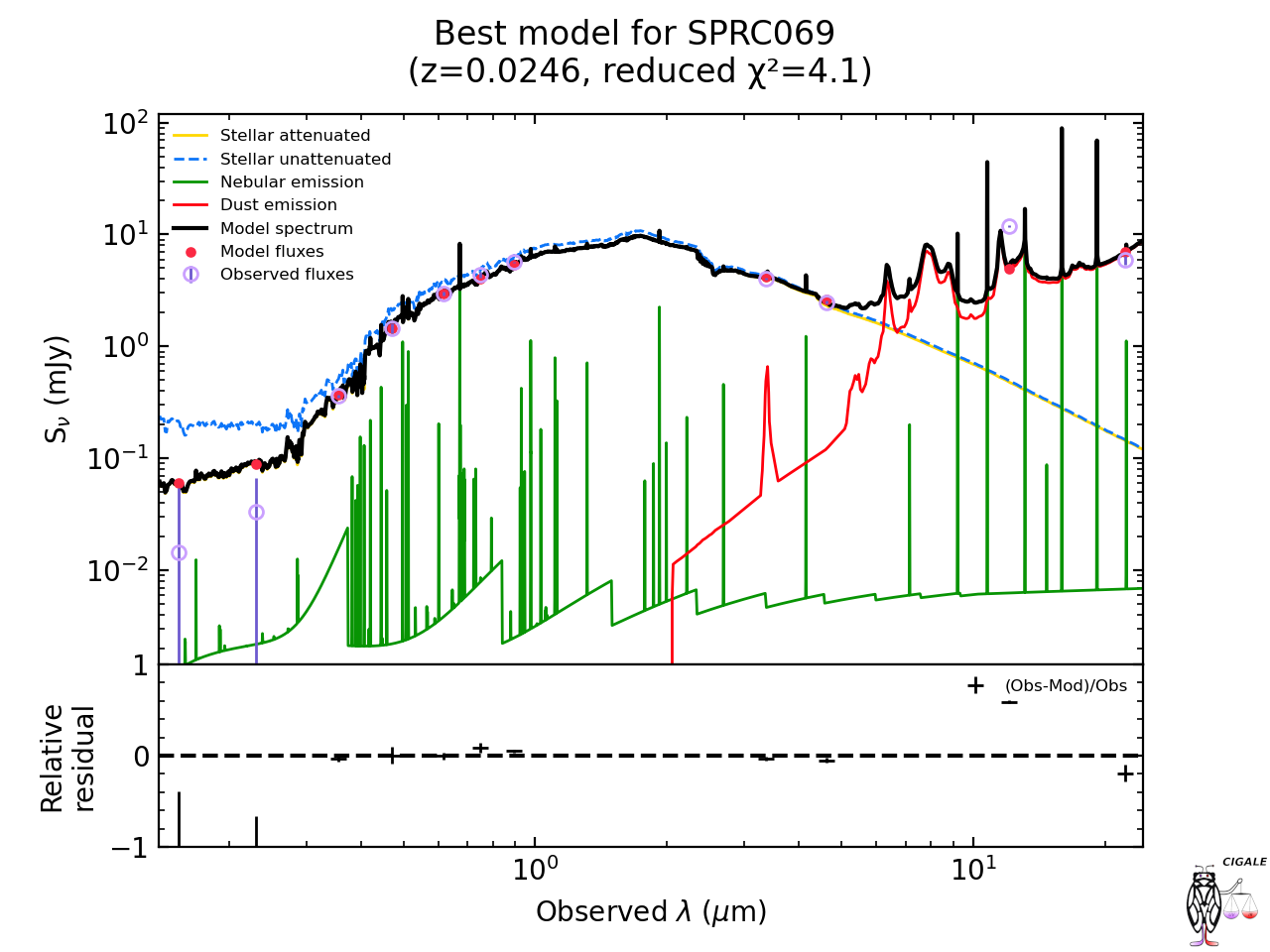}
\end{tabular} 
    \caption{Best-fit spectral SED models for the PRGs sample. Purple points show the observed fluxes, while red points mark the corresponding model fluxes. The black solid line shows the total model SED.}
    \label{fig:SED2}  
\end{figure}

\begin{figure}[!htbp]
    \ContinuedFloat
    \centering
    \begin{tabular}{@{}cc@{}}
    \includegraphics[width=0.39\textwidth]{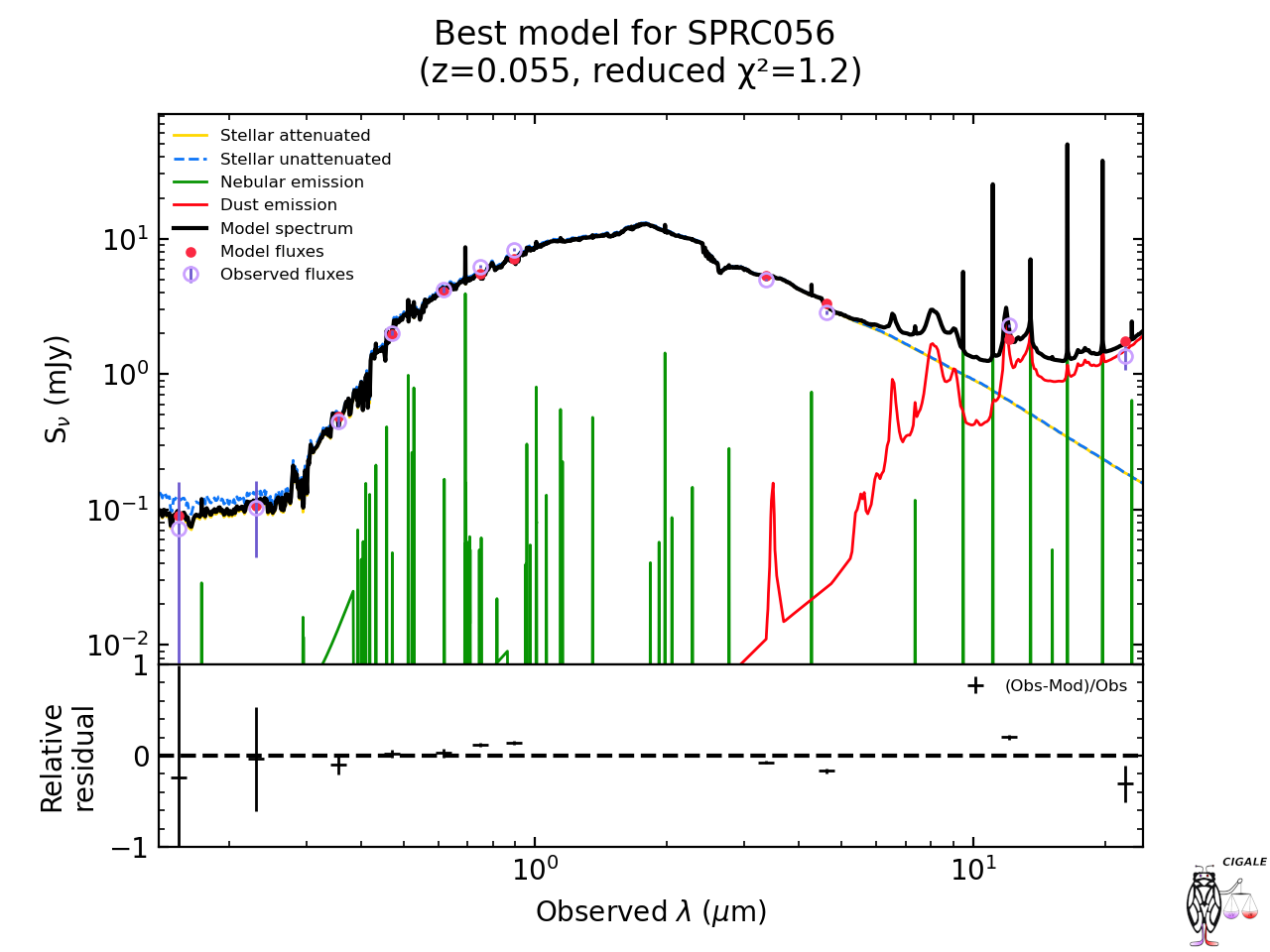} & \includegraphics[width=0.39\textwidth]{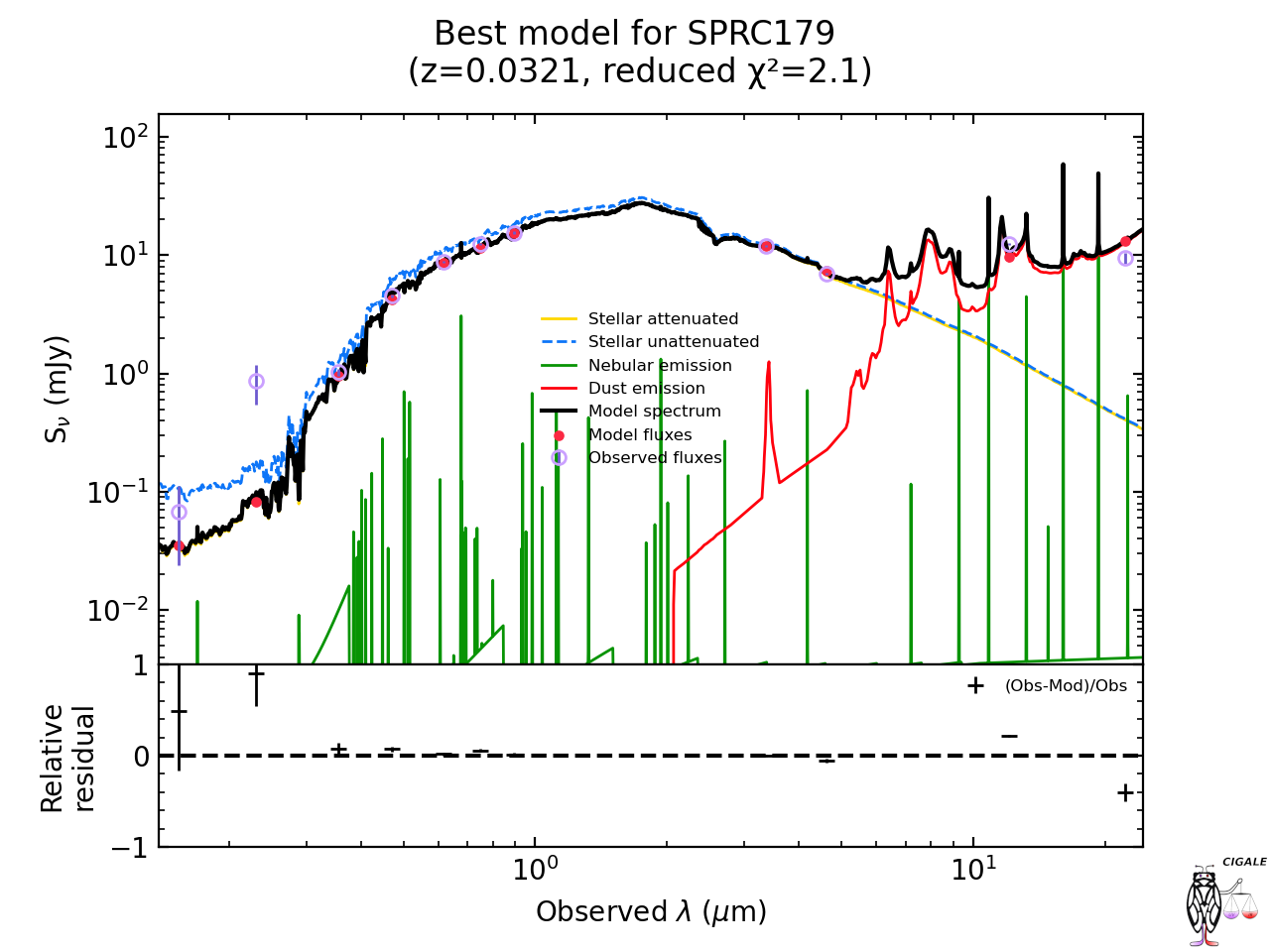} \\
    \includegraphics[width=0.39\textwidth]{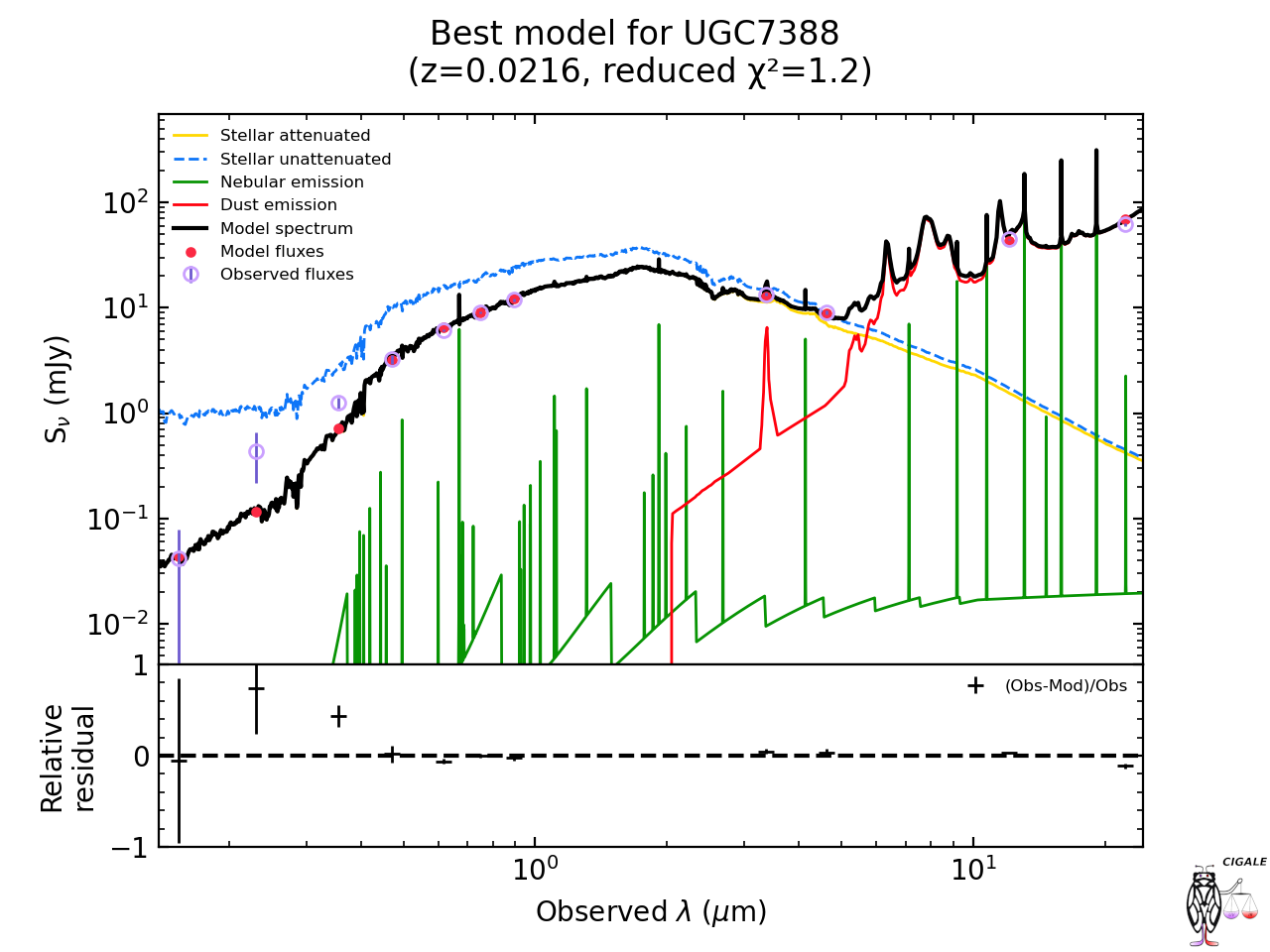} &
    \end{tabular} 
    \caption[]{Best-fit spectral energy distribution models for the PRGs sample (continued).}
\end{figure}

\twocolumn

\section{CIGALE mock analysis}
\label{app:mock}

To assess the recoverability of the physical parameters derived from
SED modeling, we performed the mock analysis implemented in CIGALE.
For each galaxy, the best-fitting model was used to generate an
artificial photometric data set by perturbing the model fluxes according
to the observational uncertainties. The resulting mock SEDs were
analyzed using the same model grid and fitting procedure as the
observed data. Comparison between the input (``exact'') and
recovered (``estimated'') values provides an empirical assessment of
parameter recovery from the available GALEX, SDSS, and WISE photometry.

Figure~\ref{fig:mock_recovery} presents the recovery tests
for the main stellar-population, SFH, and attenuation
parameters.
The present-day star-formation rate and stellar mass show
strong correlations between the input and recovered values,
with squared Pearson correlation coefficients of
$r^2=0.96$ and $r^2=0.98$, respectively.
The attenuation parameters yield
$r^2\simeq0.9998$ for both $A_{V,\rm ISM}$ and
$A_{V,\rm BC}$.
These results support the internal recoverability of the
SFR, stellar mass, and attenuation parameters for the mock
sample within the adopted model grid.
The test does not include uncertainties associated with
alternative model assumptions.
Because $\mu$ is fixed, $A_{V,\rm BC}$ and $A_{V,\rm ISM}$
are linked by the adopted attenuation prescription, and
their recovery tests are not independent.

The SFH timescales and stellar-population ages show
more variable recovery.
The main-component e-folding timescale,
$\tau_{\rm main}$, yields $r^2=0.19$, while the
secondary-component timescale, $\tau_{\rm burst}$,
yields $r^2=0.32$.
The ages of the main and secondary components yield
$r^2=0.17$ and $r^2=0.60$, respectively.
The mass-weighted stellar age shows the weakest recovery,
with $r^2\simeq0.006$.

The weak recovery of $\tau_2\equiv\tau_{\rm burst}$ is consistent with
the broad Bayesian uncertainties across the sample.
Within the adopted two-component SFH parameterization, the inferred
$\tau_2$ values characterize the secondary component on broad timescales,
with limited precision in the duration of individual star-formation
episodes. The comparison between $\tau_2$ and the Cosmic Web environment is therefore exploratory, 
with the substantial overlap between the Bayesian estimates limiting 
the interpretation of differences among individual systems.

\begin{figure}
    \includegraphics[width=0.425\textwidth]{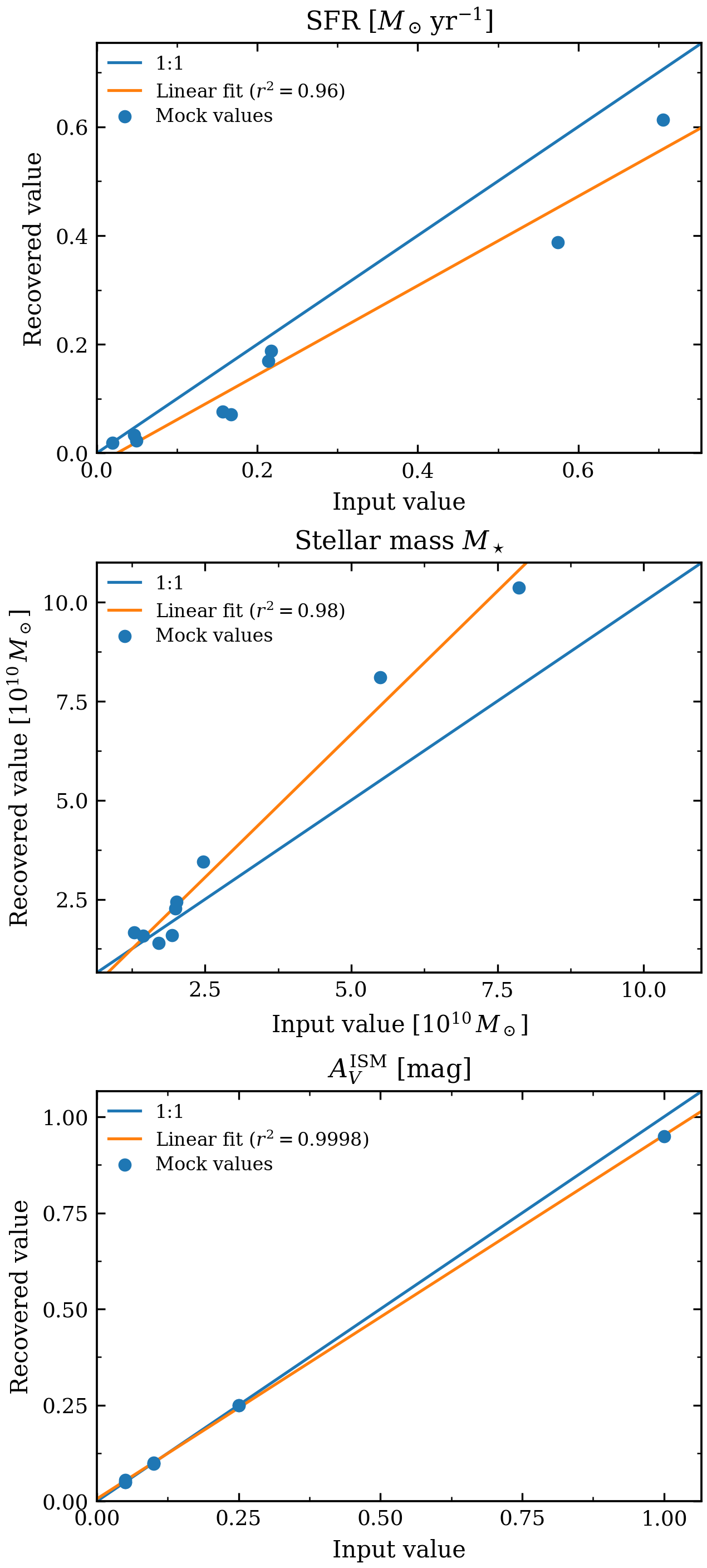}
    \caption{CIGALE mock analysis for the 
    parameters considered in the SED modeling. The panels show the
    SFR, stellar mass, attenuation in the
    diffuse ISM and birth clouds, the e-folding timescales of the main
    and secondary SFH components, the ages of the main and secondary
    stellar populations, and the mass-weighted stellar age.}
    \label{fig:mock_recovery}
\end{figure}

\begin{figure*}
\ContinuedFloat
    \includegraphics[width=0.99\textwidth]{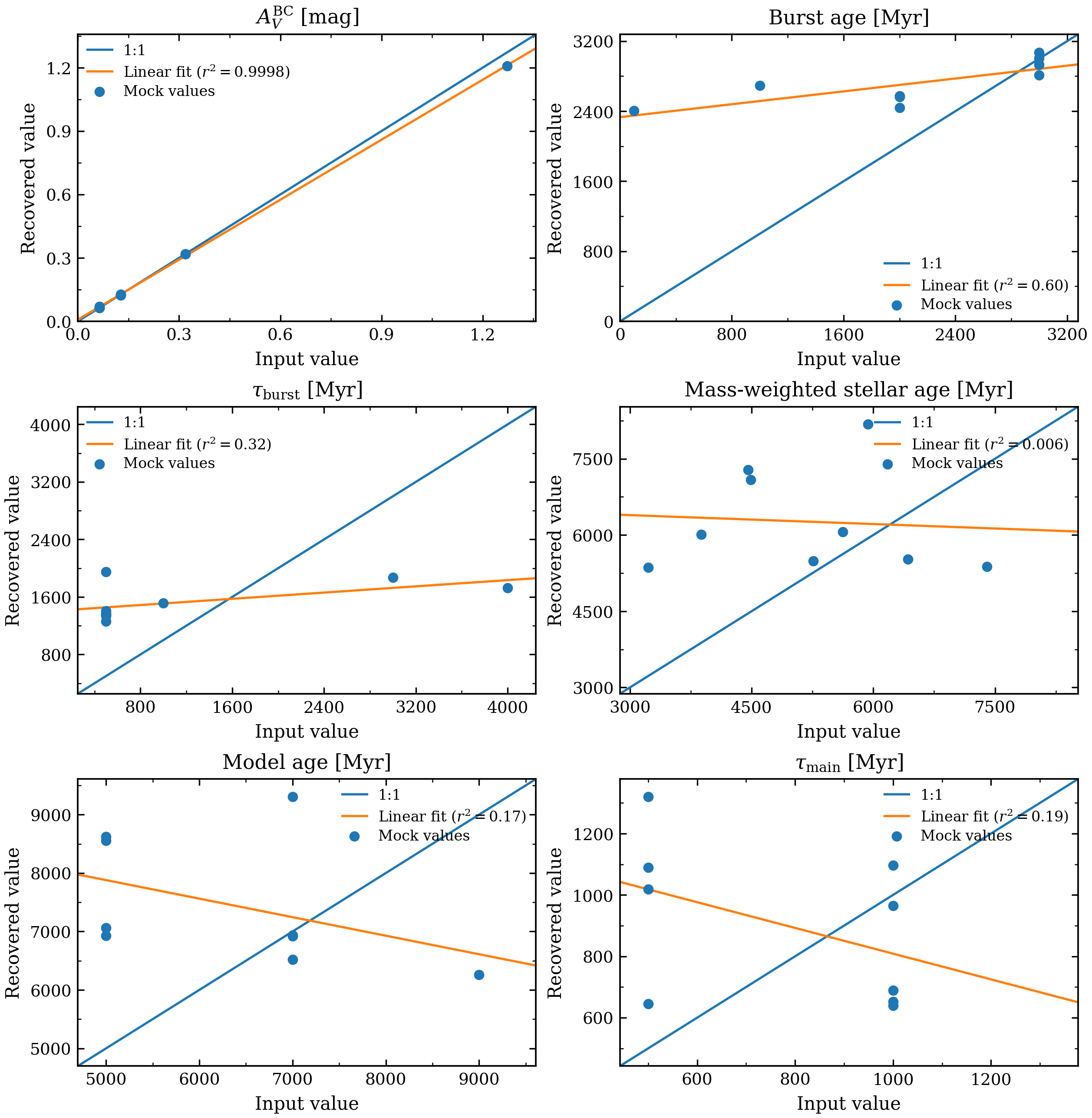} 
    \caption[]{Results of the CIGALE mock analysis for the principal
    parameters considered in the SED modeling (continued).}
\end{figure*}

\section{CIGALE Star formation histories for PRGs}
\label{app:sfh}

In this Appendix, we assess the reliability of the
star-formation-history parameters inferred from the CIGALE
SED modelling and evaluate their usefulness for the
interpretation of possible polar-ring formation scenarios. The mock analysis shows that the SFH parameters are weakly recoverable, limiting their diagnostic value. They provide supplementary qualitative constraints alongside the independent environmental information presented in the main text. Future far-infrared observations could improve the constraints on the dust-emission component of the SED, reduce modelling degeneracies, and increase the reliability of the inferred SFH parameters.

\begin{figure*}[htbp]
    \centering
    \includegraphics[width=\textwidth]{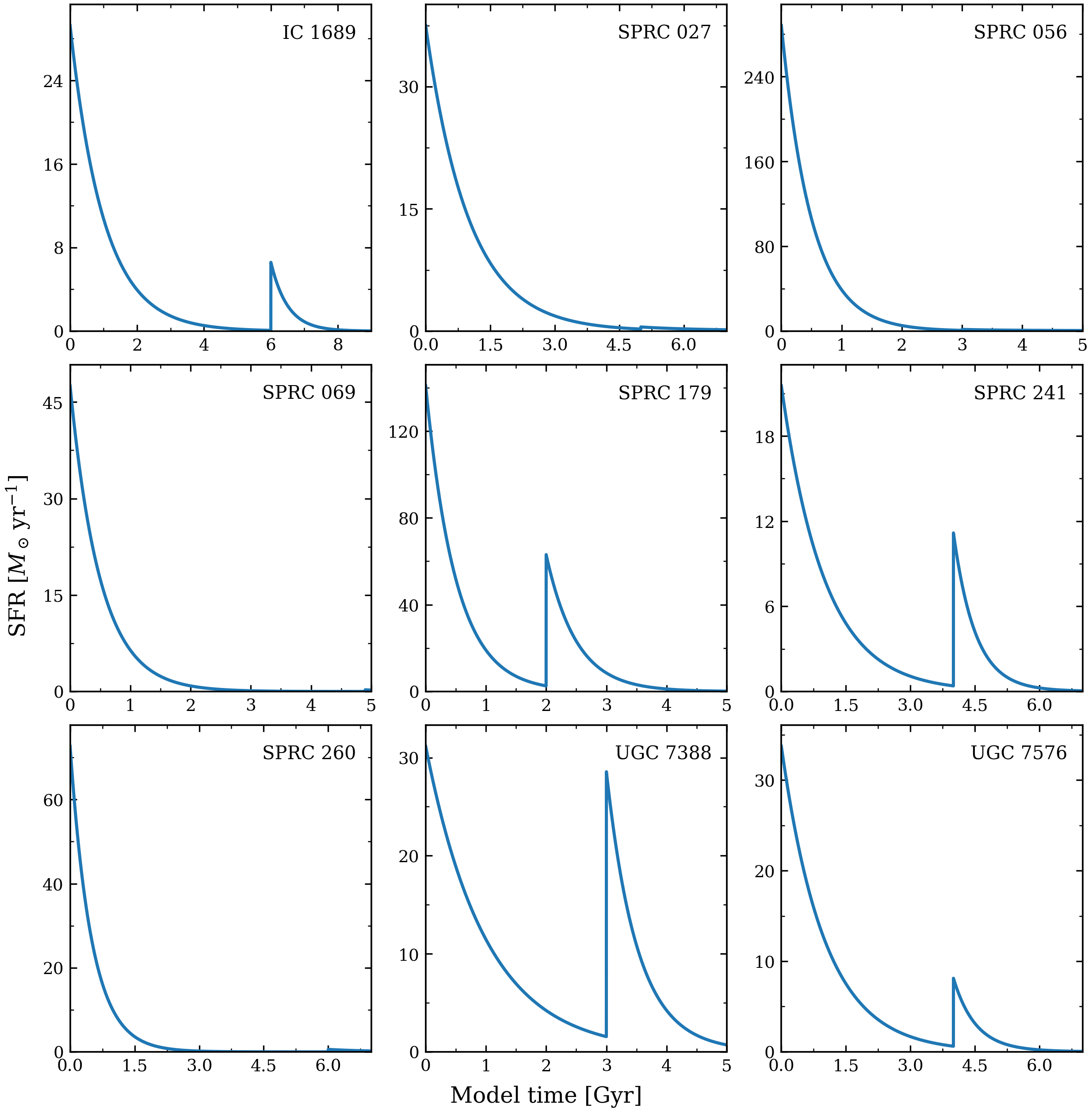}
    \caption{Star formation histories corresponding to the best-fitting CIGALE models for PRGs in sample. The time and SFR ranges are adjusted individually for each galaxy to display the shape of the reconstructed SFH.}
    \label{fig:sfh}
\end{figure*}

Figure~\ref{fig:sfh} presents the star formation histories (SFHs)
corresponding to the best-fitting CIGALE models for the galaxies in our sample. The time coordinate is shown as provided by the CIGALE SFH models, with $t=0$ corresponding to the onset of star formation in the adopted model. The different time ranges among the galaxies reflect the best-fitting model ages.

The best-fitting SFHs illustrate the temporal behaviour permitted by the adopted model family. The weak recovery of the temporal parameters limits their use in distinguishing formation pathways for individual galaxies.

\end{document}